\documentclass[prx,a4paper,aps,twocolumn,superscriptaddress,10pt,bibnotes]{revtex4-1}
\usepackage{graphicx,color}
\usepackage{dcolumn}
\usepackage{bm}
\usepackage{amsmath,amssymb,mathrsfs}
\usepackage{url}
\usepackage{hyperref}

\setcitestyle{sort&compress,numbers}

\newcommand{\set}[1]{\mathsf{#1}}
\newcommand{\grp}[1]{\mathsf{#1}}
\newcommand{\spc}[1]{\mathcal{#1}}

\def\d{{\rm d}}

\def\>{\rangle}
\def\<{\langle}
\def\kk{\>\!\>}
\def\bb{\<\!\<}

\newcommand{\map}[1]{\mathcal{#1}}
\newcommand{\Tr}{\operatorname{Tr}}

\newcommand{\op}[1]{\operatorname{#1}}

\newtheorem{theo}{Theorem}

\newtheorem{lemma}{Lemma}

\newtheorem{defi}{Definition}

\def\Proof{{\bf Proof.~}}
\def\qed{$\blacksquare$ \medskip}

\begin{document}
\title{Experimental demonstration  of  input-output indefiniteness  in a single quantum device}
    
\author{Yu Guo}
    \thanks{These two authors contributed equally.}
    \affiliation{CAS Key Laboratory of Quantum Information, University of Science and Technology of China, Hefei, 230026, China}
    \affiliation{CAS Center For Excellence in Quantum Information and Quantum Physics, University of Science and Technology of China, Hefei, 230026, China}
\affiliation{Hefei National Laboratory, University of Science and Technology of China, Hefei, 230088, China}

\author{Zixuan Liu}
    \thanks{These two authors contributed equally.}
    \affiliation{QICI Quantum Information and Computation Initiative, Department of Computer Science, The University of Hong Kong, Pokfulam Road, Hong Kong}
    \affiliation{HKU-Oxford Joint Laboratory for Quantum Information and Computation}

\author{Hao Tang}
\affiliation{CAS Key Laboratory of Quantum Information, University of Science and Technology of China, Hefei, 230026, China}
\affiliation{CAS Center For Excellence in Quantum Information and Quantum Physics, University of Science and Technology of China, Hefei, 230026, China}
\affiliation{Hefei National Laboratory, University of Science and Technology of China, Hefei, 230088, China}
    
\author{Xiao-Min Hu}
\affiliation{CAS Key Laboratory of Quantum Information, University of Science and Technology of China, Hefei, 230026, China}
\affiliation{CAS Center For Excellence in Quantum Information and Quantum Physics, University of Science and Technology of China, Hefei, 230026, China}
\affiliation{Hefei National Laboratory, University of Science and Technology of China, Hefei, 230088, China}

\author{Bi-Heng Liu}
\email{bhliu@ustc.edu.cn}
\affiliation{CAS Key Laboratory of Quantum Information, University of Science and Technology of China, Hefei, 230026, China}
\affiliation{CAS Center For Excellence in Quantum Information and Quantum Physics, University of Science and Technology of China, Hefei, 230026, China}
\affiliation{Hefei National Laboratory, University of Science and Technology of China, Hefei, 230088, China}

\author{Yun-Feng Huang}
\affiliation{CAS Key Laboratory of Quantum Information, University of Science and Technology of China, Hefei, 230026, China}
\affiliation{CAS Center For Excellence in Quantum Information and Quantum Physics, University of Science and Technology of China, Hefei, 230026, China}
\affiliation{Hefei National Laboratory, University of Science and Technology of China, Hefei, 230088, China}
    
\author{Chuan-Feng Li}
\email{cfli@ustc.edu.cn}
\affiliation{CAS Key Laboratory of Quantum Information, University of Science and Technology of China, Hefei, 230026, China}
\affiliation{CAS Center For Excellence in Quantum Information and Quantum Physics, University of Science and Technology of China, Hefei, 230026, China}
\affiliation{Hefei National Laboratory, University of Science and Technology of China, Hefei, 230088, China}

\author{Guang-Can Guo}
\affiliation{CAS Key Laboratory of Quantum Information, University of Science and Technology of China, Hefei, 230026, China}
\affiliation{CAS Center For Excellence in Quantum Information and Quantum Physics, University of Science and Technology of China, Hefei, 230026, China}
\affiliation{Hefei National Laboratory, University of Science and Technology of China, Hefei, 230088, China}

\author{Giulio Chiribella}
\email{giulio@cs.hku.hk}
\affiliation{QICI Quantum Information and Computation Initiative, Department of Computer Science, The University of Hong Kong, Pokfulam Road, Hong Kong}
\affiliation{Department of Computer Science, University of Oxford, Wolfson Building, Parks Road, Oxford, UK}
\affiliation{HKU-Oxford Joint Laboratory for Quantum Information and Computation}
\affiliation{Perimeter Institute for Theoretical Physics, 31 Caroline Street North, Waterloo,  Ontario, Canada}

\begin{abstract}
Quantum theory allows information to flow through  a single  device in a coherent superposition of two opposite directions, resulting into situations where the input-output direction is indefinite.    Here we introduce a theoretical method  to witness  input-output indefiniteness in a single quantum device, and we experimentally demonstrate it by constructing a photonic setup that exhibits input-output indefiniteness with a statistical significance exceeding 69 standard deviations. Our results provide a way to characterize input-output indefiniteness as a resource for quantum information and photonic quantum technologies and enable table-top simulations of hypothetical scenarios  exhibiting quantum indefiniteness in the direction of time.
\end{abstract}

\maketitle

\textit{Introduction.---} A cornerstone of quantum theory is the CPT theorem \cite{schwinger1951theory,luders1954equivalence}, stating that the fundamental dynamics of quantum fields is invariant under inversion of time direction, charge, and parity. The theorem implies that, at the fundamental level, the roles of past and future are symmetric: while we normally treat  systems at earlier times as the inputs  and systems at later times as the outputs, the dynamical laws of quantum mechanics are indifferent to the direction of time.   
  The time symmetry of the fundamental quantum dynamics was later extended to scenarios involving measurements by  Aharonov and collaborators \cite{aharonov1964time,reznik1995time,aharonov2010time}. With the advent of quantum information,  the role of time symmetry in quantum theory has attracted renewed attention, due to its connection with the structure of quantum protocols  \cite{abramsky2004categorical},  multitime quantum states \cite{aharonov2002two,aharonov2009multiple,silva2014pre},  simulation of closed timelike curves \cite{svetlichny2011time,lloyd2011closed,genkina2012optimal}, inversion of unknown quantum evolutions \cite{quintino2019reversing,trillo2023universal,yoshida2023reversing},  quantum retrodiction   \cite{parzygnat2023time,parzygnat2023axioms}, and the origin of irreversibility \cite{maccone2009quantum,di2021arrow}.     Time-symmetric frameworks for quantum theory \cite{oreshkov2015operational,chiribella2021symmetries} and more general physical theories \cite{hardy2021time,selby2022time} have been developed and analyzed.

Recently, Refs.\cite{chiribella2022quantum,liu2023quantum}  extended the notion of time-reversal to a broader notion of input-output inversion, which applies whenever the roles of the input and output ports of a quantum device can be exchanged. This includes, for example, the case of linear optical devices, which  can be traversed in two opposite spatial directions.  Notably, all kinds of input-output inversions turned out to share the same mathematical structure. As a consequence, hypothetical scenarios involving the reversal of the time direction between two spacetime events can be simulated by real-world setups that reverse the  direction  of a path between two points in space.
Building on the notion of input-output inversion, Ref.\cite{chiribella2022quantum} then introduced a new type of operations that utilize quantum devices in a coherent superposition of two alternative input-output directions, giving rise to a  feature called input-output indefiniteness.   This feature has been found to offer advantages in  information-theoretic \cite{chiribella2022quantum,liu2023quantum} and thermodynamical tasks \cite{rubino2021quantum,rubino2022inferring}. Input-output indefiniteness is also related to the notion of indefinite order \cite{hardy2007towards,chiribella2009beyond,oreshkov2012quantum,chiribella2013quantum}, whose applications to  quantum information have been extensively investigated in the past decade, both theoretically \cite{chiribella2012perfect,chiribella2013quantum,araujo2014computational,guerin2016exponential,ebler2018enhanced,zhao2020quantum,felce2020quantum,chiribella2021quantum} and experimentally \cite{procopio2015experimental,rubino2017experimental,goswami2018indefinite,wei2019experimental,guo2020experimental,rubino2021experimental,cao2021experimental,nie2022experimental,yin2023experimental}. An important difference is that, while  indefinite order requires multiple  devices (or multiple  uses of the same device),  input-output indefiniteness can already arise at the single-device level, enabling quantum protocols that could not be achieved  with indefinite order
(see Appendices \cite{SM} for  examples in the tasks of   gate transformation, estimation, and  testing).  



Here we develop a general method for witnessing input-output indefiniteness in the laboratory, and we use it to   experimentally demonstrate a photonic setup that probes a single quantum device  in a coherent superposition  of two alternative directions.  By optimizing the choice of witness, we demonstrate incompatibility of our setup with a definite input-output direction by more than 69 statistical deviations. Notably, our setup applies not only  to reversible quantum devices, such as polarization rotators, but also  to a  class of irreversible devices including postselected polarization measurements.  In addition to single-device indefiniteness, we experimentally demonstrate the combination of two devices in  a quantum superposition of two opposite input-output directions, building a setup that  achieves 99.6\% winning probability in a quantum  game where every  strategy using both devices in the same  direction fails with at least 11\% probability. Our techniques enable a rigorous characterization of input-output indefiniteness as a resource for quantum information and photonic quantum technologies, and, at the same time, could be used to simulate exotic  physical models where the arrow of time is subject to quantum indefiniteness.   

\begin{figure}
    \centering
    \includegraphics[width=0.4\textwidth]{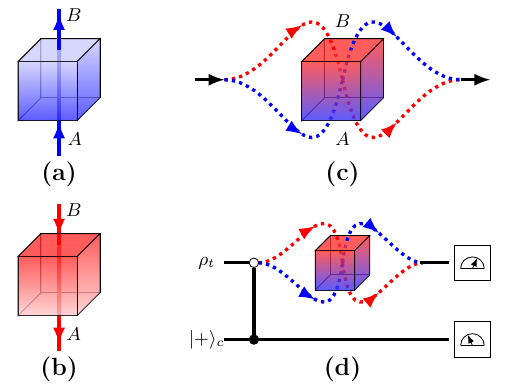} 
    \caption{{\em Input-output indefiniteness in a bidirectional quantum device.} A bidirectional device with ports $A$ and $B$ can be traversed in two opposite directions:  from $A$ to $B$  (a)  or from $B$ to $A$ (b).  When these two configurations take place in a quantum superposition (c), the direction of the information flow between  $A$ and $B$ becomes indefinite. To generate the superposition, we introduce a control qubit  that coherently controls the direction, with basis states $|0\>$ and $|1\>$ corresponding to directions $A\to B$ and $B\to A$, respectively.  
    Our  setup (d) sets the control qubit in the state $|+\>_{\rm c}  =  (|0\>+|1\>)/\sqrt 2$ and witnesses   input-output indefiniteness by performing local measurements on the target and control system,  with the target initialized in a  quantum state $\rho_{\rm t}$. }
    \label{fig:qflip}
\end{figure}


\textit{Witnesses of input-output indefiniteness.---} 
 For many processes in nature, the role of the input and output ports can be exchanged. 
  An example is  the transmission of a single photon through an optical crystal, schematically illustrated in Fig. \ref{fig:qflip}.
   Quantum devices with exchangeable input-output ports,  called {\em bidirectional},  can be used in two alternative ways,  conventionally referred to as the ``forward mode''  (with the inputs entering at port $A$ and the outputs exiting from port $B$) and ``backward mode''  (with the inputs entering at port $B$ and the outputs exiting from port $A$). In the special case where ports $A$ and $B$ are associated with two  moments of time $t_A  <  t_B$,  the forward mode corresponds to the standard use of the device in the forward time direction, while  the backward mode corresponds to a hypothetical use of the device in the reverse time direction \cite{chiribella2022quantum}.

Ref. \cite{chiribella2022quantum} showed that a  device is bidirectional if and only if the corresponding transformation of density matrices 
is a bistochastic quantum channel \cite{landau1993birkhoff,mendl2009unital}, that is, a linear map $\map C$  of the form  $\map C  (\rho)   =  \sum_i  C_i \rho  C_i^\dag$, where $\rho$ is the input density matrix, and $(C_i)$ are square matrices satisfying the conditions $\sum_i  C_i^\dag C_i  =\sum_i C_i C_i^\dag =  I$, $I$ being the identity matrix.  If a bistochastic channel $\map C$ describes the state change in the forward mode, then the state change in the backward mode is described by a (generally different) bistochastic channel $\Theta (\map C)$ given by $\Theta (\map C) :  \,  \rho  \mapsto \sum_i \theta (C_i) \, \rho\,  \theta(C_i)^\dag$, where  the square matrix $\theta (C_i)  $ is either unitarily equivalent to $C_i^T$, the transpose of $C_i$, or unitarily equivalent to $C_i^\dag$, the adjoint of $C_i$ \cite{chiribella2022quantum}.   The map $\Theta$ is called an {\em input-output inversion}.  Physically, it can represent a time reversal (if the two ports of the device correspond to two moments of time),  an inversion of spatial directions (as in the example of the optical crystal), or any other symmetry transformation obeying a set of general axioms specified in \cite{chiribella2022quantum}.   In the following, we will focus on the case where the input-output inversion is (unitarily equivalent to) the transpose.  This case  includes in particular  the canonical time-reversal in quantum mechanics  \cite{wigner1959group,messiah1965quantum} and quantum thermodynamics \cite{campisi2011colloquium}  (see \cite{SM} for more details.)

\begin{figure*}[htbp]
    \centering
    \includegraphics[width=1.3\columnwidth]{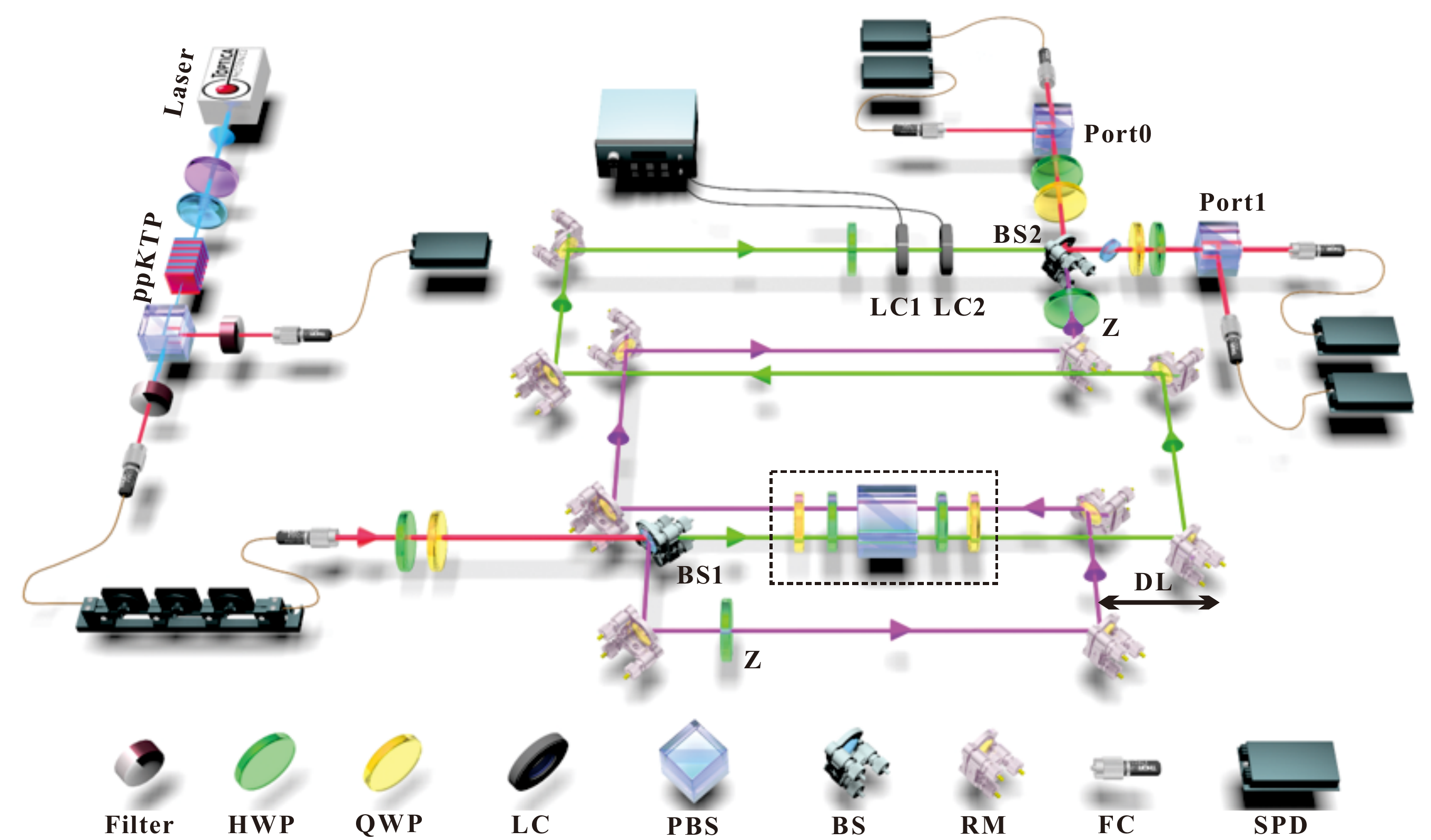}
    \caption{{\em Experimental setup.} A 2.5~mW continuous wave violet laser at 404~nm pumps a type-II cut ppKTP crystal, effectively working as a heralded single photon source when the idler photons trigger an SPD. The single photon's polarization serves as the target qubit and is initialized with a fiber polarizer controller, an HWP, and a QWP. Spatial modes of the photon serve as the control qubit, and BS1 is used to coherently control the input-output direction.  Measure-and-reprepare operations on the polarization are implemented by two HWPs, two QWPs, and a PBS (dotted rectangle), while measurements on spatial modes are implemented by two LCs and BS2. A trombone-arm delay line and a piezoelectric transducer are used to set the path length and the relative phases of the interferometer. HWP, half-wave plate; QWP, quarter-wave plate; PBS, polarizing beam splitter; BS, beam splitter; RM, reflection mirror; LC, liquid crystal variable retarder; FC, fiber coupler; SPD, single photon detector; DL, trombone-arm delay line.}
      \label{fig:experimentalsetup}
\end{figure*}

In principle, quantum mechanics allows for setups that coherently control the input-output direction, such as the setup shown in Fig. \ref{fig:qflip}(d). 
 We now develop a method for witnessing input-output indefiniteness in the laboratory. A witness for a given  quantum resource, such as entanglement \cite{GUHNE2009Entanglement}, indefinite causal order \cite{araujo2015witnessing}, and causal connection \cite{milz2022resource}, is an observable quantity that distinguishes between resourceful  and non-resourceful setups \cite{Chitambar2019Quantum}.   In our case, the non-resourceful setups are those that use the device in a well-defined direction.  Setups that use it in the forward (backward) mode are described by a suitable set of positive operators,   denoted by $\set{S}_{\rm fwd}$  ($\set{S}_{\rm bwd}$).  The explicit characterization of these operators is provided in the Appendices \cite{SM}.   For the following discussion, it will suffice  to know that they  act on the tensor product Hilbert space 
$\spc H_{A_{\rm I}} \otimes \spc H_{A_{\rm O}} \otimes \spc H_{B_{\rm I}} \otimes \spc H_{B_{\rm O}}$, where $\spc H_{A_{\rm I}}$  ($\spc H_{A_{\rm O}}$) is the Hilbert space of the input (output) system of the device, while $\spc H_{B_{\rm I}}$  ($\spc H_{B_{\rm O}}$) is the Hilbert space of the input (output) system of the overall process obtained  by inserting the  device into the setup. 

A setup that uses the device in a random mixture of the forward and backward modes  corresponds to an operator of the form 
\begin{align}\label{definite}
    S  =  p\,  S_{\rm fwd}  +  (1-p)  \,  S_{\rm bwd}\,,
    \end{align}
with $S_{\rm fwd}   \in  {\set S}_{\rm fwd}$, $S_{\rm bwd}   \in  {\set S}_{\rm bwd}$, and $p\in  [0,1]$.   We will denote by $\set{S}_{\rm definite}$ the set of all operators of the form  (\ref{definite}).   The setups  outside $\set{S}_{\rm definite}$ are incompatible with the use of the given device in a definite input-output direction: in these setups, the device is not used in the forward mode, nor in the backward mode, nor in any random mixture thereof. For an operator $S$ outside $\set{S}_{\rm definite}$,  we define a  {\rm witness of input-output indefiniteness} to be a self-adjoint operator   $W$ such that
\begin{equation}
    \Tr(WS) < 0 \, ,
\end{equation}
and
\begin{equation}
    \label{eq:witness}
 \Tr(WS') \geq 0 \, , \qquad     \forall S' \in \set{S}_{\rm definite}  \, .
\end{equation}
The condition (\ref{eq:witness}) is characterized in the following Theorem, which provides a systematic way to construct witnesses of input-output indefiniteness.

\begin{theo}
    \label{theo:witness}
    A Hermitian operator $W$ satisfies Eq. (\ref{eq:witness}) if and only if there exist operators $W_0$ and $W_1$ such that $W \geq W_0$,  $W \geq W_1$,   
    $    W_0 = {}_{[B_{\rm O}]}W_0 - {}_{[A_{\rm O}B_{\rm O}]}W_0 + {}_{[A_{\rm I}A_{\rm O}B_{\rm O}]}W_0 - {}_{[A_{\rm I}A_{\rm O}B_{\rm I}B_{\rm O}]}W_0$, 
    and
        $W_1 = {}_{[B_{\rm O}]}W_1 - {}_{[A_{\rm I}B_{\rm O}]}W_1 + {}_{[A_{\rm I}A_{\rm O}B_{\rm O}]}W_1 - {}_{[A_{\rm I}A_{\rm O}B_{\rm I}B_{\rm O}]}W_1$,  having used the notation 
   $ {}_{[X]}S := \Tr_X [S] \otimes \frac{I_X}{d_X}$
for a system $X$ of dimension $d_X$.  
\end{theo}
The proof is provided in the Appendices \cite{SM}, where we also show  that the expectation value of any witness  can be decomposed into a linear combination of outcome probabilities arising from settings in which a  device is inserted in the setup and the resulting process is probed on multiple input states. 

\textit{Experimental demonstration of input-output indefiniteness of a single quantum device.---} Our experimental setup, illustrated in Fig. \ref{fig:experimentalsetup}, is inspired by a theoretical primitive known as the quantum time flip (QTF) \cite{chiribella2022quantum}. The QTF takes in input an arbitrary bidirectional device  and adds quantum control to the direction in which the device is used. When applied to a bidirectional device that acts as channel $\map C$ in the forward direction, the QTF  generates a new quantum channel $\map F (\map C)$, acting jointly on the target system and on a control qubit.   Explicitly, the Kraus operators of the new quantum channel $\map F (\map C)$, denoted by $\{F_i\}$, are related to the Kraus operators of the original channel $\map C$, denoted by $\{C_i\}$,  as 
\begin{equation}\label{Kraus_QTF}
    F_i = C_i \otimes |0\>\<0| + C_i^T \otimes |1\>\<1|\, ,
\end{equation}
where   $\{|0\> ,|1\>\}$ are two orthogonal states of the control qubit. When the control qubit is initialized in a coherent superposition of $|0\>$ and $|1\>$,  the new channel  $\map F(\map C)$ implements a superposition of channel  $\map C$ and its input-output inversion $\map C^T$, in the sense of Refs.  \cite{aharonov1990superpositions,aaberg2004subspace,aaberg2004operations,oi2003interference,oi2003interference,chiribella2019quantum,abbott2018communication,dong2019controlled,vanrietvelde2021universal}.



\begin{figure*}[htbp]
    \centering
    \includegraphics[width=0.7\linewidth]{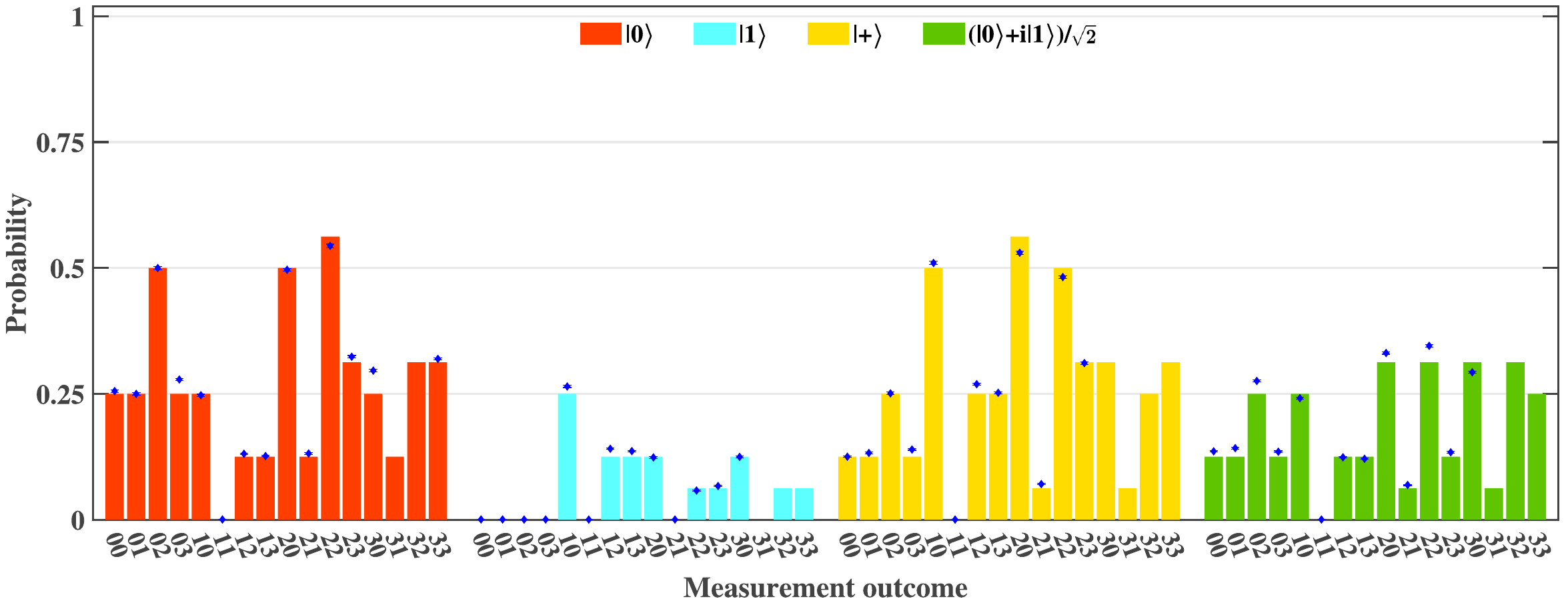}
    \caption{\emph{Experimental data for the optimal witness.} The figure shows the outcome probabilities of different measurements on the control and target qubits, with the control initialized in the state $|+\>$ and the target in one of the states  $|0\>$ (red), $|1\>$ (cyan), $|+\>$ (yellow), and $(|0\>+i|1\>)/\sqrt{2}$ (green).    The measurement outcomes are labeled by numbers 0, 1, 2, 3, corresponding to projections on  the states  $|0\>,|1\>,|+\>,(|0\>+i|1\>)/\sqrt{2}$, respectively. For example, `03' labels  the  outcome that projects the control qubit onto $|0\>$ and the target qubit onto $(|0\>+i|1\>)/\sqrt{2}$. The bars show the theoretical predictions, while the blue diamonds show the experimental data. We omit the experimental data for  outcomes  that are irrelevant to the evaluation of the optimal witness. All the data in  this figure refer to the setting where  the   device inside our setup implements a measure-and-reprepare process. Specifically, they refer to the event where the target is measured on  the $X$-eigenstate $|+\>$ and re-prepared in the $Z$-eigenstate $|0\>$. The experimental data for the remaining settings are shown in the Appendices  \cite{SM}.     }  
    \label{fig:witnessresult}
\end{figure*}

  In our experiment, schematically illustrated in Fig. \ref{fig:qflip}(d), a heralded single photon is generated through spontaneous parametric down-conversion \cite{SM}. The polarization qubit, serving as the target system in the QTF, is initialized in an arbitrary fixed state, using a fiber polarizer controller, a half-wave plate (HWP) and a quarter-wave plate (QWP). The photon is sent to a 50/50 beamsplitter (BS1) to prepare the spatial qubit in the superposition state  $|+\>  = (|0\>+|1\>)/\sqrt 2$, where $|0\>$ and $|1\>$ correspond to the two alternative paths shown green and carmine in Fig. \ref{fig:experimentalsetup}. The input device for the QTF is a bistochastic measure-and-reprepare operation \cite{chiribella2021symmetries}, implemented by an assemblage of  two HWPs, two QWPs, and a polarizing beam splitter (PBS),  shown inside  the dotted  rectangle in Fig.  \ref{fig:experimentalsetup}. The input-output inversion is realized by routing the photon through the same assemblage along a backward path sandwiched between two fixed Pauli gates $Z=|0\>\<0|-|1\>\<1|$. A coherent superposition of the forward and backward  measure-and-reprepare operations is created by using the spatial qubit as a control qubit. Finally,  two paths are coherently recombined on BS2, followed by a measurement on the polarization qubit.

To certify input-output indefiniteness, we derived the  witness $W^{\rm opt}$ with maximum robustness  to noise (see Appendices \cite{SM}.) This  witness can be estimated by probing the setup on  a set of bistochastic  measure-and-reprepare  processes that measure the polarization qubit in the eigenbasis of a Pauli gate and reprepare the output in a state in the eigenbasis of another Pauli gate. The overall evolution induced by the setup is  probed by initializing  the path qubit in the maximally coherent state $|+\>$ and   the polarization qubit  in one of the states $|0\>$,  $|1\>$, $|+\>$ and $\frac{1}{\sqrt{2}}(|0\>+i|1\>)$.  Finally,  the target qubit and control qubit are measured in the eigenbases of the three Pauli gates. The measured probabilities, shown in Fig. \ref{fig:witnessresult},  are used to calculate the experimental value of the witness $\Tr[W^{\rm opt}  \, S_{\rm QTF}]$, which we find to be $-(0.345 \pm 0.005)$,
corresponding to a violation of the condition of definite input-output direction by more than 69 standard deviations.

To implement the optimal witness $W^{\rm opt}$,  we performed  local operations on 5 qubits, using a total of 794 settings \cite{SM}.  The complexity of the experiment is less than that of a full process tomography, which would require at least 1023 settings.  To further reduce  the complexity,  we designed a simplified witness where the target qubit is initialized in a fixed  state $|0\>$ and is eventually discarded.   This witness   involves only 3 qubits and 48 settings, which we show to be the optimal values \cite{SM}.   In the experiment, we find the value  $-(0.140 \pm 0.004)$, which  certifies incompatibility with a definite input-output direction by more than 35 standard deviations. 

\textit{Experimental  demonstration of advantage in a quantum game.---}   
Input-output indefiniteness offers an advantage in a quantum game where a referee challenges a player to find out a hidden relation between two unknown quantum gates \cite{chiribella2022quantum}. In this game, the referee provides the player with two devices implementing unitary gates $U$ and $V$, respectively, promising that the two gates satisfy either the relation $U  V^T  =    U^T  V$  or the relation $U V^T =- U^T V$. The player's task is to determine which of these two alternatives holds. Ref.  \cite{chiribella2022quantum} showed that a  player that uses the two gates in the QTF  
can win the game with certainty, while every strategy that uses the two devices in the same input-output direction will fail at least 11\% of the times. 

  In our experiment, 
   discussed  in the Appendices  \cite{SM}, 
  we observe an average success probability of $99.60 \pm 0.18\%$ over a set of 21 gate pairs. The worst-case error probability  is approximately $0.68\pm 0.19\%$, which is 16 times smaller than 11\%, the lower bound on the error probability for all possible strategies with definite-input-output direction. We also show that the advantage of input-output indefiniteness persists even if  the player has coherent control on each of the gates $U$ and $V$:  every strategy using the controlled gates  $\tt{ctrl}-U  =  I\otimes |0\>\<0|  +  U\otimes |1\>\<1|$ and $\tt{ctrl}-V  =  I\otimes |0\>\<0|  +  V\otimes |1\>\<1|$   in the same input-output direction will necessarily have an error probability of at least 5.6\%.  Overall, this game can be regarded as a bipartite witness of global input-output indefiniteness.  
  In the Appendices \cite{SM}, we provide a general theory of such witnesses.

\textit{Conclusions.---} In this paper we introduced  the notion of witness of input-output indefiniteness and used it to experimentally demonstrate input-output indefiniteness in a single photonic device.    
Our results provide a way to rigorously characterize input-output indefiniteness in the laboratory, and represent a counterpart to  recent experiments on indefinite order of quantum gates \cite{procopio2015experimental,rubino2017experimental,goswami2018indefinite,wei2019experimental,guo2020experimental,rubino2021experimental,cao2021experimental,nie2022experimental,yin2023experimental}. Overall, input-output indefiniteness provides a new resource for quantum information protocols, and could potentially lead to  advantages in photonic quantum technologies.Our setup and its generalizations could also be used to simulate exotic physics in which the arrow of time is a quantum variable.   These hypothetical phenomena fit into a broad framework developed by  Hardy \cite{hardy2007towards}, who suggested that a full-fledged theory of quantum gravity would require spacetime structures  to be subject to quantum indefiniteness.  
While an explicit physical model for  scenarios with indefinite time direction has yet to be proposed, the availability of a mathematical framework for their study and  an experimental platform for their simulation   represent valuable tools for understanding their operational implications.

\begin{acknowledgements}
We acknowledge helpful discussions with Jonathan Barrett,  Hl\'er Kristj\'ansson, Kavan Modi,  Andreas Winter, Ge Bai, and Fei Meng. This work was supported by the National Key Research and Development Program of China (No.~2021YFE0113100), NSFC (No.~12374338, No.~11904357, No.~12174367,  No.~12204458, and No. 17326616), the Hong Kong Research Grant Council (No.~17300920, SRFS2021-7S02, and T45-406/23-R),  the Innovation Program for Quantum Science and Technology (No. 2021ZD0301200), the Fundamental Research Funds for the Central Universities, USTC Tang Scholarship, Science and Technological Fund of Anhui Province for Outstanding Youth (2008085J02), China Postdoctoral Science Foundation (2021M700138), China Postdoctoral for Innovative Talents (BX2021289), and the John Templeton Foundation through the ID\# 62312 grant, as part of the ‘The Quantum Information Structure of Spacetime’ Project (QISS).  This work was partially carried out at the USTC Center for Micro and Nanoscale Research and Fabrication.  The opinions expressed in this publication are those of the authors and do not necessarily reflect the views of the John Templeton Foundation.
\end{acknowledgements}

\bibliography{references}

\appendix

\section{Relation with indefinite order}\label{app:DirectionVsOrder}

Here we discuss the relation between indefinite input-output direction and indefinite order, highlighting similarities and differences, and presenting a list of quantum protocols that can be achieved using indefinite input-output direction but cannot be achieved using indefinite order.  

Indefinite order refers to scenarios where two or more quantum devices are connected with one another in a way that is not compatible with any probabilistic mixture of well-defined orders \cite{oreshkov2012quantum}.   The simplest example of this situation is the  quantum SWITCH \cite{chiribella2009beyond,chiribella2013quantum}, an operation $\map S$ that takes in input two quantum devices $\map A$ and $\map B$, acting on the same target system, and produces in output a new quantum device $\map S  (\map A, \map B)$  that adds quantum control to the order in which $\map A$ and $\map B$ are applied.  Mathematically, the new device $\map S  (\map A, \map B)$  is a  bipartite quantum channel, acting on the target system and on a control qubit that determines the relative order of $\map A$ and $\map B$.  The Kraus operators of  channel $\map S  (\map A, \map B)$, denoted by $\{S_{ij}\}$ are given by 
\begin{align}
S_{ij}   =  A_i B_j \otimes |0\>\<0|  +  B_jA_i \otimes |1\>\<1| \, ,  
\end{align}
where $\{A_i\}$ and $\{B_j\}$ are the Kraus operators of channels $\map A$ and $\map B$, respectively, and $\{|0\>, |1\>\}$ are orthogonal states of the control qubit.    When the control qubit is initialized in the state $|0\>$ ($|1\>$), the target qubit undergoes  processes $\map A$ and $\map B$ in the definite order $\map A\map B$ ($\map B\map A$).  When the control qubit is in a coherent quantum superposition of $|0\>$ and $|1\>$, instead, the order of $\map A$ and $\map B$ becomes indefinite.   
A number of information-theoretic applications of the quantum switch has been discussed over the past years \cite{chiribella2012perfect,chiribella2013quantum,araujo2014computational,guerin2016exponential,ebler2018enhanced,zhao2020quantum,felce2020quantum,chiribella2021quantum}, and series of experiments inspired by  the quantum switch has been performed on photonic systems \cite{procopio2015experimental,rubino2017experimental,goswami2018indefinite,wei2019experimental,guo2020experimental,rubino2021experimental,cao2021experimental,nie2022experimental,yin2023experimental} (see also Refs.  \cite{Paunkovic2020causalorders,vilasini2023embedding,Ormrod2023causalstructurein} for a theoretical  discussion on the interpretation of the experiments.)

Another famous example of indefinite order is a process introduced by Oreshkov, Costa, and Brukner \cite{oreshkov2012quantum}.  More generally,  the quantum switch, the Oreshkov-Costa-Brukner process, and other operations with indefinite causal order are represented by quantum supermaps \cite{chiribella2008transforming,chiribella2013quantum}, that is, higher-order maps acting on quantum channels.  This type of supermaps take in input  two (or more)  quantum channels and produce  a new quantum channel as output. 

For indefinite order, the input channels can be arbitrary completely positive, trace-preserving maps. Physically, this condition guarantees that the corresponding supermaps could in principle be implemented on arbitrary quantum devices.  In stark contrast, operations with indefinite input-output direction can only be applied to bidirectional devices, mathematically described by  bistochastic channels.  In other words, operations with indefinite input-output direction are defined on a strictly smaller domain.  The restriction of the domain leads to a broader set of conceivable supermaps, which can sometime lead to stronger advantages in quantum information tasks.   In the following, we will present three such advantages:  

\begin{enumerate}
\item {\em Unitary black box inversion/transposition.}   In this task, one is given a black box implementing an unknown quantum dynamics, represented by a unitary gate $U$ acting on a $d$-dimensional quantum system. The goal is to produce a new black box implementing the inverse of the original dynamics, corresponding to the unitary $U^\dag$, or the transpose of the original dynamics, corresponding to the gate $U^T$.   

For qubits, the inverse  and the transpose are unitarily equivalent, due to the relation  $U^\dag  =   Y  U^T  Y$, valid for every matrix $U  \in  \grp{SU}  (2)$.   For higher dimensional systems, we will focus our attention on the transpose.   If the gate $U$ is an arbitrary unitary matrix,     the transformation $U\mapsto  U^T$ cannot be perfectly achieved by inserting the gate $U$ in a quantum  circuit:  every implementation of this transformation must necessarily be probabilistic, or approximate \cite{chiribella2016optimal,quintino2019probabilistic,quintino2019reversing}.     Furthermore,  Ref. \cite{chiribella2022quantum}  showed that  the transpose $U^T$  cannot be perfectly generated  from two copies of the original gate $U$,  {\em even if the two copies are used in an indefinite order.}  

The above  no-go theorems do not apply if the experimenter is able to use the black box  in two opposite input-output directions, that is, if the black box can be treated by the experimenter as a bidirectional quantum device, acting as $U$ in one direction, and acting as $VU^TV^\dag$ in the opposite direction,  where $V$ is a fixed unitary.  In this case, the experimenter has only to use the device in the appropriate  direction and to undo the unitary $V$.   

 In summary, the transformation $U\mapsto U^T$ provides an  example of a quantum task that cannot be perfect achieved  by  operations with indefinite order  (using two queries of the unitary gate $U$),  but     can be perfectly achieved by operations with indefinite input-output direction.

\item {\em Gate estimation.}   In this task an experimenter is given  access to a black box implementing an unitary gate of the form $U_\theta  =   e^{- i\theta  H}$, where the generator $H$ has eigenvalues $\{0,1,\dots, d-1\}$ and satisfies the condition $H^T=  -  H$, while the shift parameter $\theta$ is in the range $[0,2\pi)$. The goal is to estimate the  $\theta$ with minimum error, that is,  to produce an estimate $\hat \theta$ that minimizes the root mean square error (RMSE)  $\Delta \theta  :  =  \sqrt{\int_{0}^{2\pi}  \d \hat \theta \,   (\hat \theta - \theta)^2  \,  p(\hat \theta | \theta)}$, where $ p(\hat \theta|\theta)$ is the conditional probability of obtaining estimate $\hat \theta$ when the true value is $\theta$.  

Suppose that the $N$ copies of the gate $U_\theta$ are available to the experimenter.  When the $N$ gates are used in parallel, they act as a single $N$-partite gate, and the total generator has spectrum $\{0,\dots, N(d-1)\}$.  If is then known that the  minimum RMSE scales as  $\Delta \theta \approx  \pi/[\sqrt  2  \, Nd]$ at the leading order in $Nd$, both in the worst case over $\theta$ and on average over uniformly distributed $\theta$ \cite{buvzek1999optimal,buvzek1999optimalmanipulations,chiribella2012optimal,gorecki2020pi}.      This value of the RMSE is optimal even if the $N$ gates are used in a sequence \cite{chiribella2008memory},  or, more generally, in an indefinite order (by a simple generalization of the argument in \cite{chiribella2008memory}).  

In contrast, the QTF can transform each gate $U_\theta$ into the gate $ W_\theta  =  U_\theta  \otimes |0\>\<0|  +  U_{\theta}^T \otimes |1\>\<1|$,  whose generator $K:  =  H \otimes Z$ has spectrum $\{-d+1,  \dots,  d-1\}$.  By applying the QTF to the initial $N$ gates, the experimenter can obtain $N$ copies of the gate $W_\theta$, which can then be used to estimate $\theta$ with 
 RMSE  $\pi /[2 \, \sqrt 2   Nd]$ at the leading order in $Nd$.      In other words, indefinite input-output direction can reduce the RMSE by a factor 2, equivalent to doubling the number of available uses of the unknown gate $U_\theta$.   

\item {\em Testing properties of quantum gates.}  
Another advantage of input-output indefiniteness  arises in the game described in the main text, originally introduced in Ref.  \cite{chiribella2022quantum}. In this game, a referee prepares a pair of black boxes implementing unitary gates $U$ and $V$, respectively, and guarantees that either  the relation $U  V^T  =    U^T  V$  or the relation $U V^T =- U^T V$ is satisfied.  A player can query each black box one time, and then has to determine which of the two alternative relations holds. 

Ref.  \cite{chiribella2022quantum} showed every strategy that uses the two devices in the same input-output direction will necessarily have a nonzero probability of error.   This conclusion applies even if the two black boxes are used in an indefinite order: if the input-output direction is fixed and equal for both boxes, then the error is non-zero.   The nature of this advantage will be  discussed in detail in Section \ref{app:game_witness}, where we classify different types of witnesses of input-output indefiniteness for pairs of bistochastic channels.

\end{enumerate}

\section{Relation with the notion of time reversal in quantum theory and quantum thermodynamics }

Here we briefly summarize  the relation between the notion of input-output inversion and the notion of  time-reversal  in quantum mechanics \cite{wigner1959group,messiah1965quantum}  and in  quantum thermodynamics \cite{campisi2011colloquium}. 

Ref. \cite{chiribella2022quantum} showed that the input-output inversion of a unitary channel, corresponding to a unitary matrix $U$, is another unitary channel, corresponding to another unitary matrix $\theta (U)$.  The map $\theta :  U \mapsto\theta (U)$  is either unitarily equivalent to the transpose  ($\theta (U)  =   V U^T V^\dag \, \forall U$, where $V$ is a fixed unitary matrix) or unitarily equivalent to the adjoint ($\theta (U)  =   V U^T V^\dag \, \forall U$, where $V$ is a fixed unitary matrix).

An example of input-output inversion where $\theta (U)$ is unitarily equivalent to the transpose arises from the  classic  notion of time reversal in quantum mechanics \cite{wigner1959group}. In this formulation, time-reversal corresponds to  a symmetry  of the state space. By Wigner's theorem, state space symmetries are described  by operators that are either unitary or anti-unitary (see e.g.  \cite{Uhl16}).  For the time-reversal symmetry, the canonical choice is to take a  anti-unitary operator, motivated by physical considerations such as the preservation of the canonical commutation relations  under the transformation $X \mapsto  X$, $P \mapsto  -  P$ \cite{messiah1965quantum}, or the requirement that the energy be bounded from below both in the forward-time picture and in the backward-time picture \cite{weinberg1995quantum,roberts2017three}.

The choice of a time reversal symmetry at the state level induces a notion of time reversal of unitary dynamics. Suppose that the time reversal symmetry is described by  an  operator $A$ (either unitary or anti-unitary). Then, if a forward-time  unitary dynamics $U$ transforms the state  $|\psi\>$ into the state $|\psi'\>=  U|\psi\>$, then the corresponding backward-time dynamics should  transform the state  $A|\psi'\>$ into  the state $A|\psi\>$,   for every possible initial state $|\psi\>$.   This condition implies that the backward-time  dynamics is also unitary, and the corresponding unitary operator $U_{\rm rev}$ satisfies   the condition
    \begin{align}
   \label{timerev}
  U_{\rm rev}    =   A   U^\dag A^{-1} \, ,
  \end{align}
  where $A^{-1}$ is the inverse of $A$.  
 This equation is known in quantum control and quantum thermodynamics, where it corresponds to the so-called {\em microreversibility principle} in the special case of autonomous ({\em i.e.} non-driven) systems with Hamiltonian invariant under time-reversal (cf. Eq. (40) of \cite{campisi2011colloquium}).

  In the canonical case where  $A$ is an anti-unitary operation, one can write $A  =   V  K  $, where $V$ is a unitary operator, and $K:  |\psi\> \mapsto  |\overline \psi\>$ is the complex conjugation in a given basis \cite{Uhl16}.    Using this decomposition, we can write the time-reversed unitary as $U_{\rm rev}  =   V  U^T  V^\dag$,  
  where $U^T$ denotes the transpose of $U$ in the given basis.

One can also consider non-canonical choices of time-reversal, such as the one advocated by Albert \cite{albert2000time} and Callender \cite{callender2000time}, who argued that, in certain systems, time-reversal should leave quantum states unchanged.   This choice yields $U_{\rm rev}  =  U^\dag$.   More generally, if one chooses   the operator $A$ to be a generic unitary $V$,   the time-reversed dynamics has the form $U_{\rm rev}  =   V  U^\dag  V^\dag$. 

It is important to stress that   input-output inversion is a more general notion than time reversal (T).  For example, it  applies to arbitrary  combinations of the time-reversal symmetry with other symmetries, such as parity inversion (P) and charge-conjugation (C).    In other words,  all the combinations CT, PT, and CPT are possible input-output inversions. 
 Regarding the full CPT symmetry, Ref. \cite{skotiniotis2013quantum} argued that it corresponds to a unitary transformation  $V$ at the state space level.  In this case, the analogue of Eq.  (\ref{timerev}) yields implies that the CPT symmetry transforms a unitary  dynamics $U$ into another unitary dynamics of the form $V U^\dag V^\dag$.

In the non-unitary case,  notions of time-reversal have been proposed  in the literature.   An early formulation is due to Crooks \cite{crooks2008quantum},  who defined  the time-reversal of a quantum channel $\map C$ as the  Petz' recovery map $\map C_{\rm Petz}$~\cite{petz1988sufficiency}, explicitly given by    $\map C_{\rm Petz} (\rho)  : =   \rho_0^{1/2}\,   \map C^\dag\,  (\rho_0^{-1/2} \, \rho \, \rho_0^{-1/2}) \, \rho_0^{1/2}$  where  $\rho_0$ is any quantum state such that $\map C (\rho_0)  =  \rho_0$, and $\map C^\dag$ is the adjoint of channel $\map C$.  If one restricts the time-reversal to  bistochastic channels and one picks the state $\rho_0$ to be maximally mixed, then Crooks' definition coincides with input-output inversion discussed in the main text, in the special case where the input-output inversion is unitarily equivalent to the adjoint.

An extension of Crook's approach  was recently proposed by  Chiribella, Aurell, and \.Zyczkowski   \cite{chiribella2020symmetries}. In this extension,  one defines   a fixed  reference state for every system, and defines the time-reversal  on the subset of channels $\map C$ satisfying the condition $\map C (\rho_{S_1})  =  \rho_{S_2}$, where $\rho_{S_1}$ and $\rho_{S_2}$ are the fixed reference states of the systems $S_1$ and $S_2$ corresponding to the input and output of channel $\map C$, respectively.  On this subset of channels, the time-reversal is defined as the Petz recovery map  $\map C_{\rm Petz} (\rho)  : =   \rho_{S_1}^{1/2}\,   \map C^\dag\,  (\rho_{S_2}^{-1/2} \, \rho \, \rho_{S_2}^{-1/2}) \, \rho_{S_1}^{1/2}$, or as the  variant  of the Petz recovery map where the adjoint $\map C^\dag$ is  replaced by the transpose $\map C^T$ (explicitly, $\widetilde{\map C}_{\rm Petz} (\rho)  : =   \overline \rho_{S_1}^{1/2}\,   \map C^\dag\,  (\overline \rho_{S_2}^{-1/2} \, \rho \, \overline \rho_{S_2}^{-1/2}) \, \overline \rho_{S_1}^{1/2}$, where $\overline \rho$ denotes the complex conjugate of the matrix $\rho$ in the given basis).  When the reference states are set to be maximally mixed, this notion of time reversal mathematically coincides with the notion of input output inversion discussed in the main text.

\section{Characterization of the witnesses}\label{app:witness}
In this section, we provide the characterization of the witnesses of input-output indefiniteness, which is done with the Choi representation \cite{choi1975completely}. The Choi representation associates  linear maps $\map M :  L(\spc H_{\rm in}) \to   L(\spc H_{\rm out})$  to bipartite operators $\op{Choi}(\map M) := (\map I_{\rm in} \otimes \map M)(|I\kk\bb I|_{\rm in, in})$,  where $L(\spc H)$ denotes the linear operators on a Hilbert space $\spc H$, and $|I\kk_{\rm in, in} = \sum_m |m\>|m\>$ is the (unnormalized) maximally entangled quantum state on $\spc H_{\rm in}^{\otimes 2}$. The Choi  representation is related to the Jamio\l kowski representation \cite{jamiolkowski1972linear} $\op{Jam}  (\map M)  : =\sum_{m,n}  |m\>\<n|  \otimes \map M  (|n\>\<m|)$ by a partial transposition on the first Hilbert space. 

In the Choi representation, a setup that uses the original device in the forward mode corresponds to a positive operator $S_{\rm fwd}$ satisfying the conditions  $\Tr_{ B_{\rm O} }  [  S_{\rm fwd}]    =   I_{ A_{\rm O}}  \otimes \Tr_{A_{\rm O}  B_{\rm O}}[S_{\rm fwd}]/ d_{A} $ and   $\Tr_{A_{\rm O} A_{\rm I} B_{\rm O}}[S_{\rm fwd}] / d_{A}    =  I_{B_{\rm I}}$ \cite{chiribella2009theoretical}. Similarly, a setup that uses the original device in the backward mode corresponds to a positive operator $S_{\rm bwd}$  satisfying the conditions $\Tr_{ B_{\rm O} }  [  S_{\rm bwd}]    =   I_{A_{\rm I}}  \otimes \Tr_{A_{\rm I}  B_{\rm O}}[S_{\rm bwd}]/ d_A $ and   $\Tr_{A_{\rm O} A_{\rm I} B_{\rm O}}[S_{\rm bwd}] / d_{A}    =  I_{B_{\rm I}}$.  More generally, a setup that uses the device in a random mixture of the forward and backward mode corresponds to an operator of the form $S  =  p\,  S_{\rm fwd}  +  (1-p)  \,  S_{\rm bwd}\,$, which, as we have mentioned in the main text, is incompatible with the use of the given device in an indefinite input-output direction. The scenarios involving indefinite input-output direction can be described by quantum supermaps \cite{chiribella2022quantum}. Every quantum supermap $\map S: \map M \mapsto  \map S (\map M)$   induces a linear  map $\widehat{\map S}$ on the Choi operators via the relation    $\widehat{\map S}  (\op{Choi}(\map M))   =   \op{Choi}(\map S(\map M))$.  Hence,  we can define the Choi operator of the supermap $\map S$  as  the Choi operator of the induced map $\widehat{S}$. In turn, the Choi operator of a supermap  acting on a single bidirectional device as in Fig. 1 of the main text can be described by a positive operator $S$, acting on the tensor product Hilbert space $\spc H_{A_{\rm I}} \otimes \spc H_{A_{\rm O}} \otimes \spc H_{B_{\rm I}} \otimes \spc H_{B_{\rm O}}$, where $\spc H_{A_{\rm I}}$  ($\spc H_{A_{\rm O}}$) is  the Hilbert space of the input (output) system of the initial channel $\map C$, while $\spc H_{B_{\rm I}}$  ($\spc H_{B_{\rm O}}$) is the Hilbert space of the input (output) system of the final channel $\map S  (\map C)$.   Since the original device $\map C$ transforms a given quantum system into itself, the Hilbert spaces $\spc H_{A_{\rm I}}$ and $\spc H_{A_{\rm O}}$ have the same dimension, hereafter denoted by $d_A$. 

In the following, the set of all witnesses of input-output indefiniteness will be denoted by $\set W_{\rm in/out}$, which will also be characterized in the Choi representation.  With this notation, we are ready to provide our characterization of the set $\set W_{\rm in/out}$.

\subsection{Proof of Theorem 1 of the main text}

Since all witnesses have non-negative expectation values on the setups with definite input-output direction,   $\set W_{\rm in/out}$ is the dual cone of $\set{S}_{\rm definite}$.  In turn, $\set{S}_{\rm definite}$ is the convex hull of the (Choi operators of) supermaps with forward input-output direction and of those with backward input-output direction.  The Choi operators of supermaps with forward input-output direction generate  a closed convex cone of operators on $\spc H_{A_{\rm I}} \otimes \spc H_{A_{\rm O}} \otimes \spc H_{B_{\rm I}} \otimes \spc H_{B_{\rm O}}$. Specifically, the closed convex cone is equal to $\spc P \cap \spc L_f$, where $\spc P$ is the cone of positive operators, $\spc L_f$ is the subspace defined by the intersection of two subspaces.
\begin{align}
    \label{eq:lfdefi}
    \spc L_f &:= \left\{ X \mid \Tr_{B_{\rm O}}  (X) = \Tr_{A_{\rm O}B_{\rm O}}  (X) \otimes \frac{I_{A_{\rm O}}}{d_{A}} \right\} \\
    & \quad \cap \left\{ X \mid \Tr_{A_{\rm I}A_{\rm O}B_{\rm O}}(X) = \frac{ \Tr (X)}{d_{B_{\rm I}}}   \cdot I_{B_{\rm I}} \right\} \, .
\end{align}
Notice that the projections onto the two subspaces in Eq. (\ref{eq:lfdefi}) commute. It follows that $\spc L_f$ can be expressed, with the notation of Theorem 1 in the main text, as below,
\begin{align}
    \label{eq:lfproj}
    \spc L_f = &\Big\{ X \mid {}_{[B_{\rm O}]}X - {}_{[A_{\rm O}B_{\rm O}]}X  + {}_{[A_{\rm I}A_{\rm O}B_{\rm O}]}X \\
    &\quad - {}_{[A_{\rm I}A_{\rm O}B_{\rm I}B_{\rm O}]}X = 0 \Big\} \, .
\end{align}
Similarly, the Choi operators of supermaps with backward input-output direction generate the closed convex cone $\spc P \cap \spc L_b$, with $\spc L_b$ given by
\begin{align}
    \label{eq:lbproj}
    \spc L_b = &\Big\{ Y \mid {}_{[B_{\rm O}]}Y - {}_{[A_{\rm I}B_{\rm O}]}Y + {}_{[A_{\rm I}A_{\rm O}B_{\rm O}]}Y \\
    &\quad - {}_{[A_{\rm I}A_{\rm O}B_{\rm I}B_{\rm O}]}Y = 0 \Big\} \, .
\end{align}
It follows that the conic hull of $\set{S}_{\rm definite}$ is equal to
\begin{equation}
    \label{eq:coni_dcircuit}
        \op{coni}(\set{S}_{\rm definite}) 
        = \op{conv}[(\spc P \cap \spc L_f) \cup (\spc P \cap \spc L_b)] \, .
\end{equation}
The dual cone of $\set{S}_{\rm definite}$ can be deduced using the duality properties of closed convex cones:
\begin{equation}
    \label{eq:witness_sets}
    \begin{split}
        &\op{coni}(\set{S}_{\rm definite})^* \\
        &= (\spc P \cap \spc L_f)^* \cap (\spc P \cap L_b)^* \, , \\
        &= \op{conv}(\spc P \cup \spc L_f^\perp) \cap \op{conv}(\spc P \cup \spc L_b^\perp) \, , \\
        &= (\spc P + \spc L_f^\perp) \cap (\spc P + \spc L_b^\perp) \, ,
    \end{split}
\end{equation}
where $\perp$ denotes orthogonal complement and $+$ denotes Minkowski addition.

Now we can conclude from Eq. (\ref{eq:witness_sets}) that a Hermitian operator $W$ on $\spc H_{A_{\rm I}} \otimes \spc H_{A_{\rm O}} \otimes \spc H_{B_{\rm I}} \otimes \spc H_{B_{\rm O}}$ is a witness of input-output indefiniteness if and only if 
\begin{equation}
    W \geq W_0 \, , \quad W \geq W_1 \, ,
\end{equation}
for some $W_0 \in \spc L_f^\perp$ and $W_1 \in \spc L_b^\perp$. According to the characterization of $\spc L_f$ and $\spc L_b$ in Eq. (\ref{eq:lfproj}) and (\ref{eq:lbproj}), respectively, $W_0$ satisfies the condition
\begin{equation}
    W_0 = {}_{[B_{\rm O}]}W_0 - {}_{[A_{\rm O}B_{\rm O}]}W_0 + {}_{[A_{\rm I}A_{\rm O}B_{\rm O}]}W_0 - {}_{[A_{\rm I}A_{\rm O}B_{\rm I}B_{\rm O}]}W_0 \, ,
\end{equation}
and $W_1$ satisfies the condition
\begin{equation}
    W_1 = {}_{[B_{\rm O}]}W_1 - {}_{[A_{\rm I}B_{\rm O}]}W_1 + {}_{[A_{\rm I}A_{\rm O}B_{\rm O}]}W_1 - {}_{[A_{\rm I}A_{\rm O}B_{\rm I}B_{\rm O}]}W_1 \, .
\end{equation}
\qed

\subsection{Measures of input-output indefiniteness}

Witnesses  can be used not only  to detect resources, but also  to  define quantitative resource measures. This is done by assessing the robustness of a given resource to the addition of noise,  as it was done, e.g. for the robustness of entanglement \cite{steiner2003generalized}, robustness of indefinite  causal order \cite{araujo2015witnessing}, and robustness of causal connection \cite{milz2022resource}. 

Setups that use  bidirectional channels are mathematically described by quantum supermaps  that transform bistochastic channels into (generally non-bistochastic) channels \cite{chiribella2022quantum}. The Choi  operator of every  such supermap, denoted by $S$, satisfies the conditions $\Tr_{A_{\rm O}A_{\rm I}B_{\rm O}}[S]/d_A = I_{B_{\rm I}}$ and
\begin{multline}
    \Tr_{B_{\rm O}}[S] = 
    \Tr_{B_{\rm O}A_{\rm O}}[S] \otimes \frac{I_{A_{\rm O}}}{d_A} \\
    + \Tr_{B_{\rm O}A_{\rm I}}[S] \otimes \frac{I_{A_{\rm I}}}{d_A} \\
    - \Tr_{B_{\rm O}A_{\rm O}A_{\rm I}}[S] \otimes \frac{I_{A_{\rm O}}}{d_A} \otimes \frac{I_{A_{\rm I}}}{d_A} \, . \label{eq:generalsupermap}
\end{multline}
In the following, the set of all Choi operators satisfying the above constraints will be denoted by $\set{S}$. 

We define the robustness of a setup $S  \in  \set{S}$ with respect to a witness $W$ as
\begin{equation}
    \label{eq:robustness_against_witness}
    r(S \mid W) := \min_{T \in \set{S}} \left\{ \lambda \geq 0 \mid \Tr\left(W \cdot \frac{S + \lambda T}{1+\lambda}\right) \geq 0 \right\} \, ,
\end{equation}
which is equal to the amount of  noise the setup $S$ can tolerate until its input-output indefiniteness stops to be detected by the witness $W$. The definition implies that the input-output indefiniteness of the setup $S$ can be detected by the witness $W$ only if the quantity $r(S \mid W)$ is larger than zero.  Optimizing the quantity $r(S \mid W)$ over the witness $W$, we obtain the robustness of input-output indefiniteness of the setup $S$ 
\begin{equation}
    \label{eq:robustness}
    r(S) := \max_{W \in \set{W}_{\rm in/out}} r(S \mid W) \, .
\end{equation}

Both Eq. (\ref{eq:robustness_against_witness}) and Eq. (\ref{eq:robustness}) can be phrased as semidefinite programming (SDP) problems and can be computed efficiently (see Section \ref{app:sdp}). In particular, the quantity $r(S)$ is given by the following SDP:
\begin{alignat}{2}
    \label{eq:robustness_sdp}
    & \text{maximize} \quad && -\Tr(WS) \\
    & \text{subject to} \quad && W \in \set{S}_{\rm definite}^* \nonumber \\
    &  \quad && \frac I {d_Ad_{B_{\rm I}}} - W \in \set{S}^* \nonumber \, ,
\end{alignat}
where $\set{S}_{\rm definite}^*$ and $\set S^*$ are the dual cones of $\set{S}_{\rm definite}$ and $\set S$, respectively. The last constraint of Eq. (\ref{eq:robustness_sdp}) can be interpreted as a normalization condition, noticing that, for every setup $T$, the value $\Tr(WT)$  should not exceed than 1. 

The form of Eq. (\ref{eq:robustness_sdp}) implies that the robustness is a convex in its argument, is faithful measure of input-output indefiniteness ($r(S)$ is equal to zero if and only if $S$ is compatible with a well-defined input-output direction), and cannot be increased by composing the setup $S$ with local bistochatic channels.


\section{Efficient computation of the robustness of input-output indefiniteness via SDP}\label{app:sdp}


\subsection{Derivation of the SDP problems}

The evaluation of the robustness of input-output indefiniteness can be cast into an SDP problem that can be solved efficiently, similarly to the SDP problem for the causal robustness in Ref. \cite{araujo2015witnessing}.   Explicitly, the SDP for the robustness of input-output indefiniteness is 
\begin{alignat}{2}
    \label{eq:primal_sdp1}
    & \text{minimize} \quad && \frac{\Tr(T)}{d_{B_{\rm I}}d_A} \\
    & \text{subject to} \quad && \Tr[W(S + T)] \geq 0 \nonumber \\
    &  \quad && T \in \op{coni}(\set{S}) \nonumber \, ,
\end{alignat}
where $\op{coni}  (\set{S})$ denotes the conic hull of the set $\set S$, consisting of the Choi operators of all deterministic supermaps on bistochastic channels.  The above expression follows from the definition of robustness in Eq. (\ref{eq:robustness_against_witness}).

The dual  of the above SDP is
\begin{alignat}{2}
    \label{eq:dual_sdp1}
    & \text{maximize} \quad && -y\Tr(WS) \\
    & \quad && \frac{I}{d_{B_{\rm I}}d_A} - yW \in \op{coni}(\set{S})^* \nonumber \\
    & \quad && y \geq 0 \, . \nonumber
\end{alignat}
In the next subsection, we  show that  the primal-dual pair (\ref{eq:primal_sdp1}) and (\ref{eq:dual_sdp1}) satisfies the condition for strong duality, meaning that  the solutions of the primal and dual problems coincide.  Hence, the robustness $r(S \mid W)$ can be computed through the dual problem (\ref{eq:dual_sdp1}).

Furthermore, the maximum robustness over all possible witnesses, denoted by  $r(S) : =  \max_S  r(S \mid W)$, can also be computed by an SDP, which follows from Eq. (\ref{eq:dual_sdp1})) by absorbing the variable $y$ into the variable $W$, thus obtaining 
\begin{alignat}{2}
    \label{eq:dual_sdp2}
    & \text{maximize} \quad && -\Tr(WS) \\
    & \quad && W \in \op{coni}(\set{S}_{\rm definite})^* \nonumber \\
    & \quad && \frac{I}{d_{B_{\rm I}}d_A} - W \in \op{coni}(\set{S})^* \nonumber \, .
\end{alignat}
The problem (\ref{eq:dual_sdp2}) is exactly the SDP problem in Eq. (\ref{eq:robustness_sdp}). Its dual problem is
\begin{alignat}{2}
    \label{eq:primal_sdp2}
    & \text{minimize} \quad && \frac{\Tr(T)}{d_{B_{\rm I}}d_A} \\
    & \text{subject to} \quad && S + T \in \op{coni}(\set{S}_{\rm definite}) \nonumber \\
    &  \quad && T \in \op{coni}(\set{S}) \nonumber \, .
\end{alignat}

\subsection{Proof of strong duality and efficient solvability}  

Here we prove that the SDP problems derived in the previous subsection can be solved efficiently and satisfy the condition for strong duality.   To obtain this result, we will use some general facts about conic optimization problems, a large category of optimization problems that include SDP as a special case.

Mathematically, a conic optimization problem is defined as follows: 
\begin{defi}\label{defi:conic}
Let $E$ be a finite-dimensional vector space, $\spc K$ a closed convex pointed cone in $E$ with a nonempty interior, and $\spc L$ a linear subspace of $E$. Let also $b \in E$ and $c \in E$. The data $E$, $\spc K$, $\spc L$, $b$, and $c$ define a pair of conic problems
\begin{align*}
(P):& \quad \textup{minimize} \ \<c, x\> \quad \textup{subject to} \quad x \in \spc K \cap (\spc L+b), \\
(D):& \quad \textup{minimize} \ \<y, b\> \quad \textup{subject to} \quad y \in \spc K^* \cap (\spc L^\perp+c), 
\end{align*}
where $\spc K^* \subseteq E$ is the cone dual to $\spc K$, $\spc L^\perp \subseteq E$ is the orthogonal complement to $\spc L$, $\spc L+b \subseteq E$ and $\spc L^\perp+c \subseteq E$ are affine subspaces.
(P) and (D) are called, respectively, the primal and dual problems associated with the above data.
\end{defi}

A relation between the optimal solutions of the primal and dual problems was provided in Theorem 4.2.1 of Ref. \cite{nesterov1994interior}: 
\begin{theo}\label{theo:duality}
Let (P), (D) be a primal-dual pair of conic problems as defined above, and let the pair be such that
\begin{enumerate}
    \item The set of primal solutions $\spc K \cap (\spc L+b)$ intersects $\op{int} \spc K$;
    \item The set of dual solutions $\spc K^* \cap (\spc L^\perp+c)$ intersects $\op{int} \spc K^*$;
    \item $\<c, x\>$ is lower bounded for all $x \in \spc K \cap (\spc L + b)$.
\end{enumerate}

Then both the primal and the dual problems are solvable with polynomial-time interior-point methods, and the optimal solutions $x^*$ and $y^*$ satisfy the relation
\begin{equation}\label{eq:duality}
\<c, b\> = \<c, x^*\> + \<y^*, b\>.
\end{equation}
\end{theo}

We now apply the above results to the primal-dual SDP pairs (\ref{eq:primal_sdp1}, \ref{eq:dual_sdp1}) and (\ref{eq:primal_sdp2}, \ref{eq:dual_sdp2}).  Let $\spc L_s$ be the linear space spanned by the operators of general supermaps on bistochastic channels, which is characterized by Eq. (\ref{eq:generalsupermap}),
\begin{align}
    \spc L_s := \Big\{ 
    &S \mid {}_{[B_{\rm O}]}S - 
    {}_{[A_{\rm O}B_{\rm O}]}S 
    - {}_{[A_{\rm I}B_{\rm O}]}S \nonumber \\
    &\quad + 2{}_{[A_{\rm O}A_{\rm I}B_{\rm O}]}S - {}_{[A_{\rm O}A_{\rm I}B_{\rm I}B_{\rm O}]}S = 0
    \Big\} \, ,
\end{align}
having used the notation
\begin{equation}\label{normalizedtracenotation}
    {}_{[X]}S := \Tr_X [S] \otimes \frac{I_X}{d_X}
\end{equation}  
for a system $X$ of dimension $d_X$. Recall the subspaces $\spc L_f$ and $\spc L_b$ defined in Section (\ref{app:witness}) which are the linear span of the operators of forward and backward setups, respectively. Since every operator of supermap we consider is in $\spc L_s$, it suffices to optimize witnesses within $\spc L_s$.
Let us start by translating these SDPs into  the language of Definition \ref{defi:conic}. To this purpose, we define the following data of a conic optimization problem:
\begin{equation}
\begin{split}
    E &= \spc L_s \times \spc L_s \, , \\
    \spc L &= \{ (T, T) \mid T \in \spc L_s \} \, , \\
    b &= (S, 0) \, , \\
    c &= \left(0, \frac I {d_{B_{\rm I}}d_A} \right) \, .
\end{split}
\end{equation}
 
\begin{figure*}[ht!]
    \centering
    \includegraphics[width=0.9\linewidth]{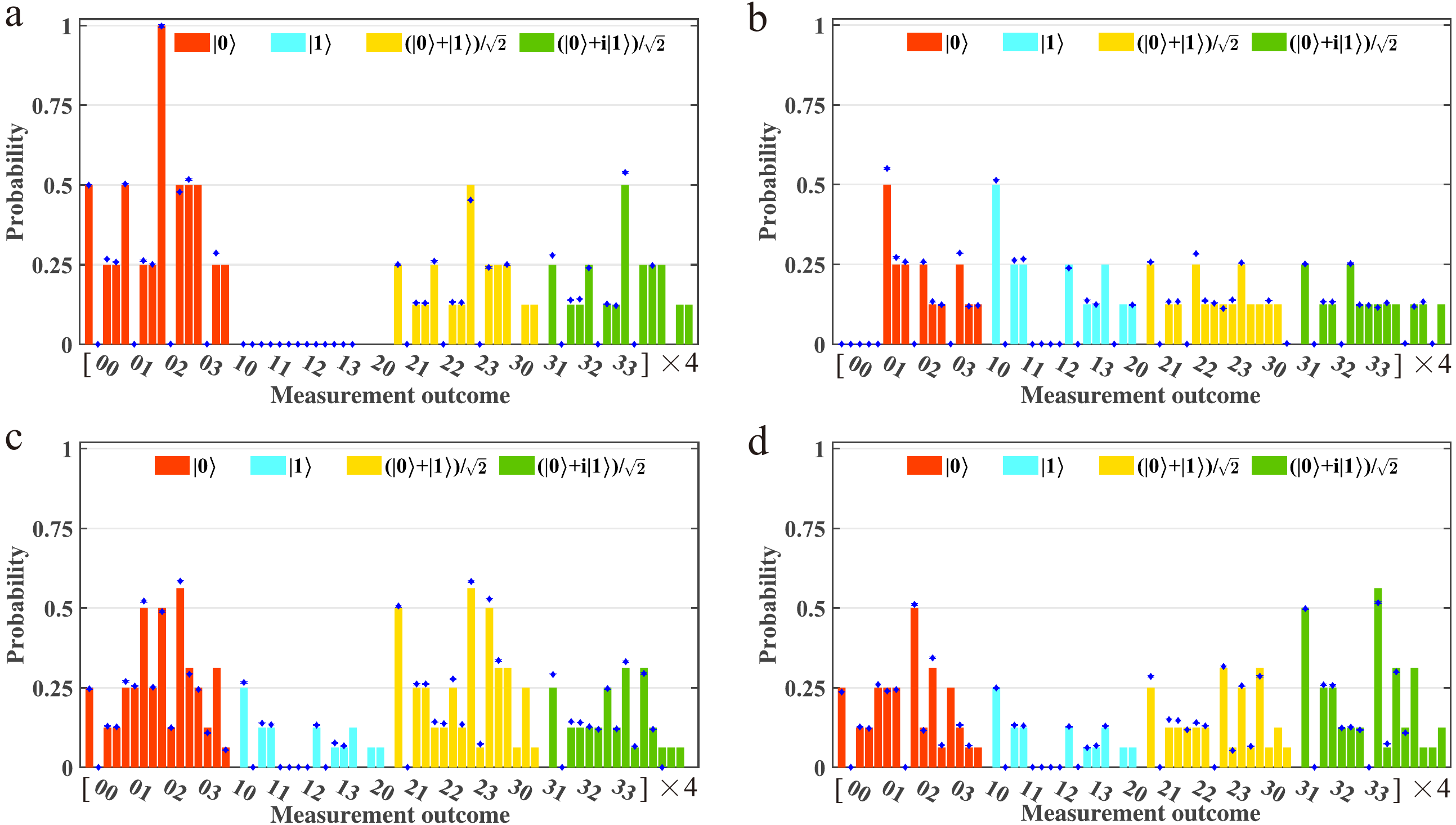}
    \caption{\emph{Experimental data for the optimal witness  (settings $|0\>\<0|$,  $|1\>\<0|$,  $|+\>\<0|$, and $|+i\>\<0|$).}    
      The figure shows the outcome probabilities of different measurements on the control and target qubits, with the control initialized  in the state $|+\>$ and the target  in one of the states  $|0\>$ (red), $|1\>$ (cyan), $|+\>$ (yellow), and $|+i\> $ (green).    The measurement outcomes are labeled by numbers 0, 1, 2, 3, corresponding to projections on  the states  $|0\>, |1\>, |+\>, |+i\>$, respectively. The bars   show the theoretical predictions, while the blue diamonds show the experimental data. We omit the experimental data for  measurement outcomes that are irrelevant to the estimation    of the  optimal witness.  The data in the figure have been taken in different settings,  corresponding to  different events of measure-and-prepare processes. 
      Specifically, Subfigures {\bf a},   {\bf b},  {\bf c}, and {\bf d} refer to the events where the target qubit is measured on the state $|0\>$ and re-prepared in the state 
      $|0\>$ ({\bf a}), $|1\>$ ({\bf b}), $|+\>$ ({\bf c}),  and 
      $|+i\>$ ({\bf d}).     }\label{fig:smfig1}
\end{figure*}

\begin{figure*}[ht!]
    \centering
    \includegraphics[width=0.9\linewidth]{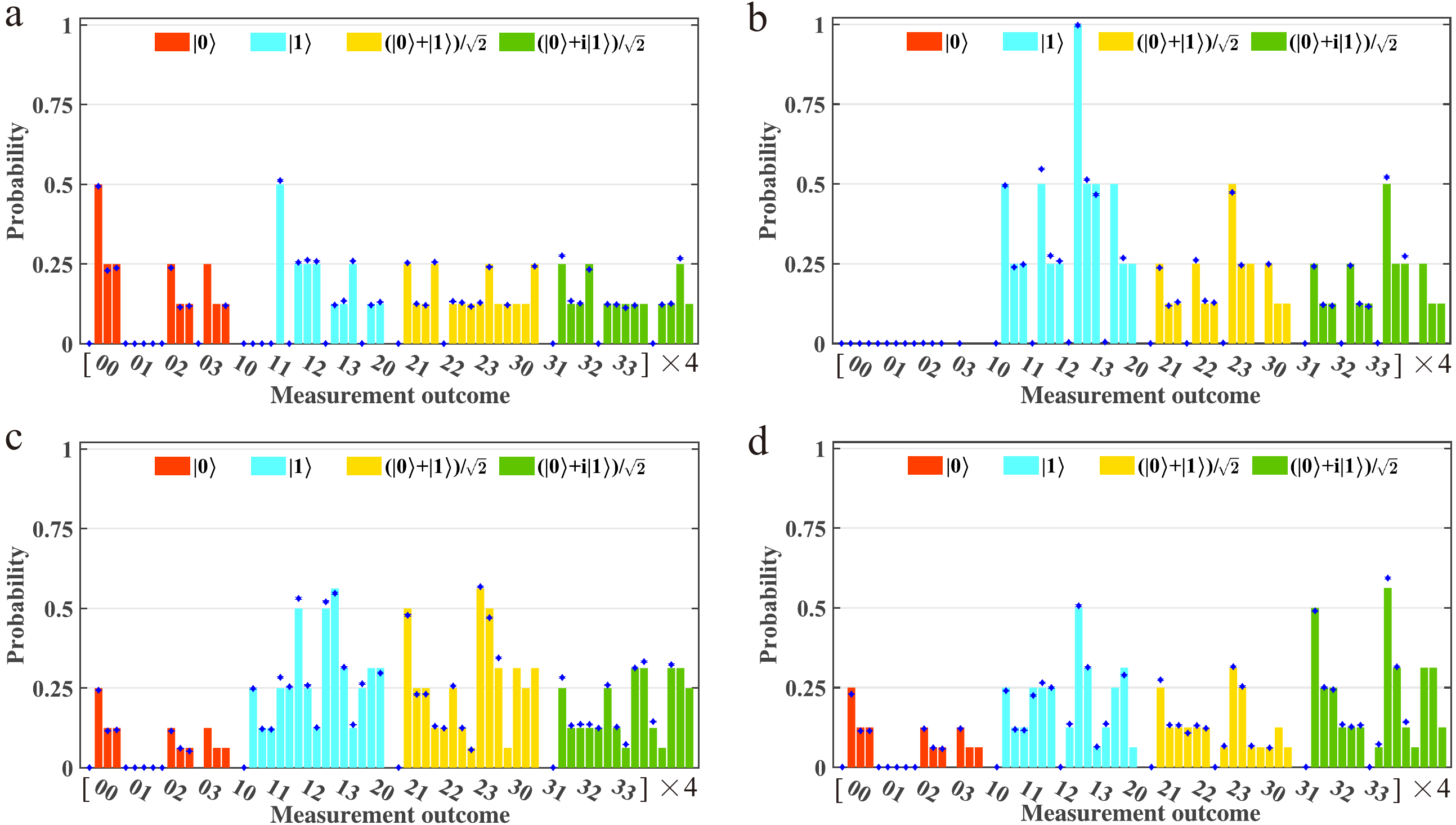}
    \caption{\emph{Experimental data for the optimal witness  (settings $|0\>\<1|$,  $|1\>\<1|$,  $|+\>\<1|$, and $|+i\>\<1|$).}    
      The figure shows the outcome probabilities of different measurements on the control and target qubits, in the setting where  the target qubit is measured on the state $|1\>$ and re-prepared in the state  
      $|0\>$ ({\bf a}), $|1\>$ ({\bf b}), $|+\>$ ({\bf c}),  and 
      $|+i\>$ ({\bf d}).   The data representation in this figure follows the same convention as in Fig. \ref{fig:smfig1}.
    }\label{fig:smfig2}
\end{figure*}

For the pair (\ref{eq:primal_sdp1}, \ref{eq:dual_sdp1}), we define
\begin{multline}
    \spc K_1 = \{ T \in \op{coni}(\set{S}) \mid \Tr(WT) \geq 0 \} \times \op{coni}(\set{S}) \, .
\end{multline}

For the pair (\ref{eq:primal_sdp2}, \ref{eq:dual_sdp2}), we define
\begin{equation}
    \spc K_2 = \op{coni}(\set{S}_{\rm definite}) \times \op{coni}(\set{S}) \, .
\end{equation}

To show that the set of primal solutions of the conic optimization problem intersects the interior of the convex cone $\spc K_1 (\spc K_2)$, we prove the following lemma which shows that the sum  of the identity operator and an arbitrary operator with a constrained norm from $\spc L_s$ is contained in the convex cone of $\set{S}_{\rm definite}$:
\begin{lemma}
\label{lemma:inner}
 Every operator $S \in \spc L_s$ with Hilbert-Schmidt norm bounded as $\Vert S \Vert_2 \leq 1/2$ satisfies the condition $I + S \in \op{coni}(\set{S}_{\rm definite})$.
\end{lemma}

\Proof This proof is similar to the proof of Lemma 7 of \cite{araujo2015witnessing}. Let $S$ be such an operator. $I + S$ can be decomposed into $S_f + S_b$ where
\begin{equation}
    S_f = \frac I 2 + {}_{[A_{\rm O}]}S \, , \quad S_b = \frac I 2 + S - {}_{[A_{\rm O}]}S \, .
\end{equation}
We can check that $S_f \in \spc L_f$ and $S_b \in \spc L_b$. Notice that ${}_{[A_{\rm O}]}S$ and $S - {}_{[A_{\rm O}]}S$ are orthogonal. By Pythagoras' theorem, it holds that
\begin{equation}
    \Vert {}_{[A_{\rm O}]}S \Vert_2^2 + \Vert S - {}_{[A_{\rm O}]}S \Vert_2^2 = \Vert S \Vert_2^2 \, .
\end{equation}
Therefore, 
\begin{multline}
    \label{eq:singularvalue}
    \max \{ \Vert {}_{[A_{\rm O}]}S \Vert, \Vert S - {}_{[A_{\rm O}]}S \Vert \} \\
    \leq \max \{ \Vert {}_{[A_{\rm O}]}S \Vert_2, \Vert S - {}_{[A_{\rm O}]}S \Vert_2 \} \leq \Vert S \Vert_2 \leq \frac 1 2 \, ,
\end{multline}
where $\Vert \cdot \Vert$ is the standard operator norm (i.e. the maximal singular value). Therefore, both $S_f$ and $S_b$ are positive. So, $S_f \in \spc P \cap \spc L_f$ and $S_b \in \spc P \cap \spc L_b$. It follows that $I + S$ belongs to $\op{coni}(\set{S}_{\rm definite})$.
\qed

Now we check that the three conditions of Theorem \ref{theo:duality} are satisfied by  the two pairs of SDP problems (\ref{eq:primal_sdp1}, \ref{eq:dual_sdp1}) and (\ref{eq:primal_sdp2}, \ref{eq:dual_sdp2}). Let $S$ be the Choi operator of a setup.
\begin{enumerate}
    \item 
    Since $\Vert S \Vert_2 \leq \Vert S \Vert_1 = \Tr(S) = d_{B_{\rm I}}d_A$ and $S \in \spc L_s$, Lemma \ref{lemma:inner} implies that
    \begin{equation}
        S + \lambda I \in \op{coni}(\set{S}_{\rm definite}) \, , \quad \forall \lambda \geq 2d_{B_{\rm I}}d_A \, .
    \end{equation}
    So by choosing $\lambda_0 > 2d_{B_{\rm I}}d_A$, we have $(\lambda_0 I, \lambda_0 I) \in \spc L$ and
    \begin{equation}
        b + (\lambda_0 I, \lambda_0 I) = (S + \lambda_0 I, \lambda_0 I) \in \op{int} \spc K_2 \, .
    \end{equation}
    
    \item Let $W_0 = \frac I {2d_{B_{\rm I}}d_A}$.
    We have $(W_0, -W_0) \in \spc L^\perp$ and
    \begin{equation}
        (W_0, -W_0) + c = (W_0, W_0) \, .
    \end{equation}
    $(W_0, W_0)$ is an interior point of the dual cone of $\set{S} \times \set{S}$ because $\Tr(W_0T) = \Tr(W_0T') = 1/2 > 0$ for every $T, T' \in \set{S}$.
    
    \item Both (\ref{eq:primal_sdp1}) and (\ref{eq:primal_sdp2}) are lower bounded by 0 because $T$ is a positive operator.
\end{enumerate}
Due to the relation
\begin{equation}
    \spc K_2 \subseteq \spc K_1 \subseteq \op{coni}(\set{S}) \times \op{coni}(\set{S}) \, ,
\end{equation}
we have
\begin{equation}
    \op{int}(\spc K_2) \subseteq \op{int}(\spc K_1) \, ,
\end{equation}
and
\begin{equation}
    \op{int}((\op{coni}(\set{S}) \times \op{coni}(\set{S}))^*) \subseteq \op{int}(\spc K_1^*) \subseteq \op{int}(\spc K_2^*) \, .
\end{equation}
It follows that the two pairs of SDP problems (\ref{eq:primal_sdp1}, \ref{eq:dual_sdp1}) and (\ref{eq:primal_sdp2}, \ref{eq:dual_sdp2}) satisfy the three conditions of Theorem \ref{theo:duality} and thus can be solved efficiently.

\begin{figure*}[ht!]
    \centering
    \includegraphics[width=0.9\linewidth]{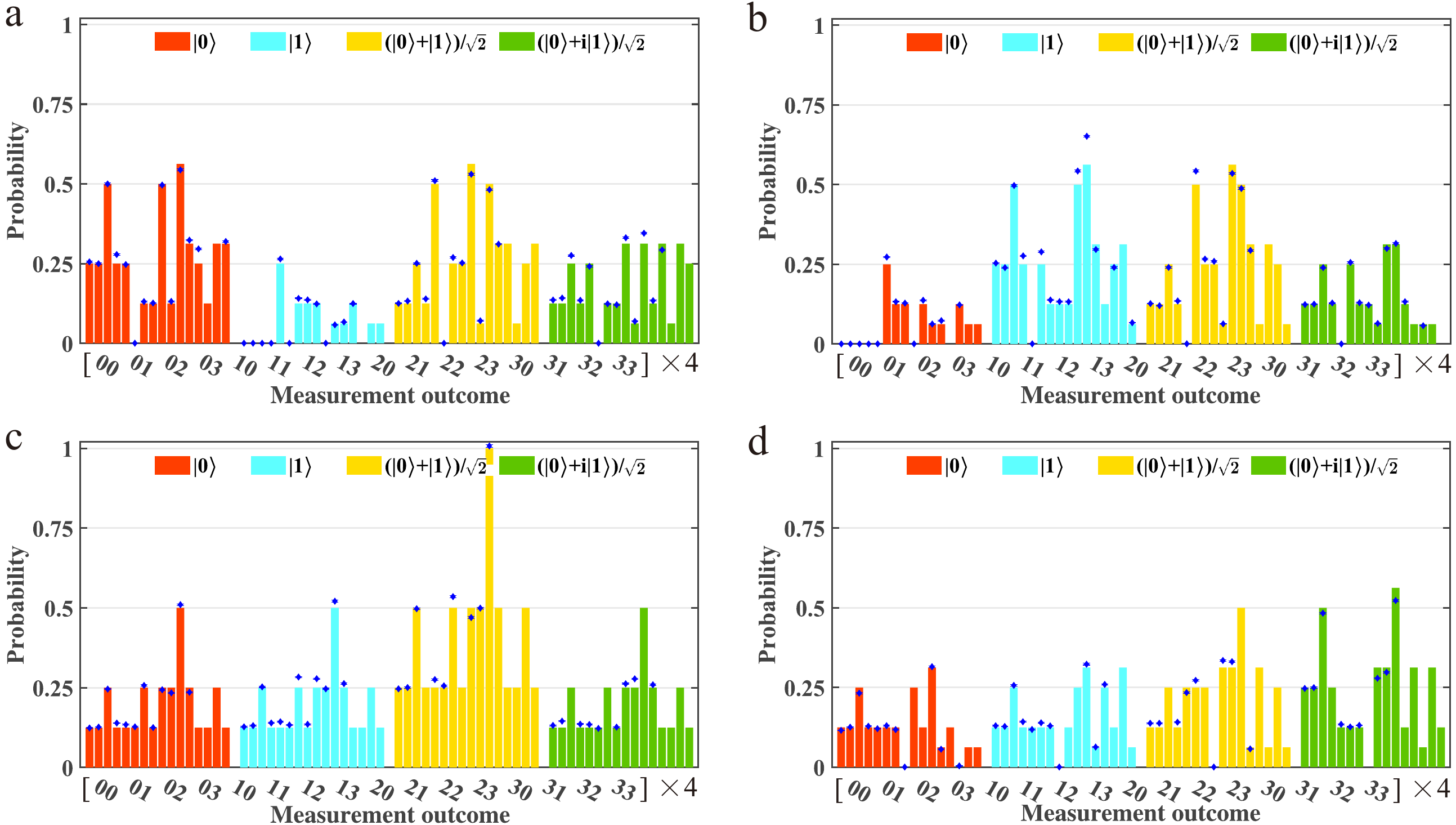}
    \caption{ \emph{Experimental data for the optimal witness  (settings $|0\>\<+|$,  $|1\>\<+|$,  $|+\>\<+|$, and  $|+i\>\<+|$).}    
      The figure shows the outcome probabilities of different measurements on the control and target qubits, in the setting where  the target qubit is measured on the state $|+\>$ and re-prepared in the state  
      $|0\>$ ({\bf a}), $|1\>$ ({\bf b}), $|+\>$ ({\bf c}),  and 
      $|+i\>$ ({\bf d}).   The data representation in this figure follows the same convention as in Fig. \ref{fig:smfig1}.  }\label{fig:smfig3}
\end{figure*}

\begin{figure*}[ht!]
    \centering
    \includegraphics[width=0.9\linewidth]{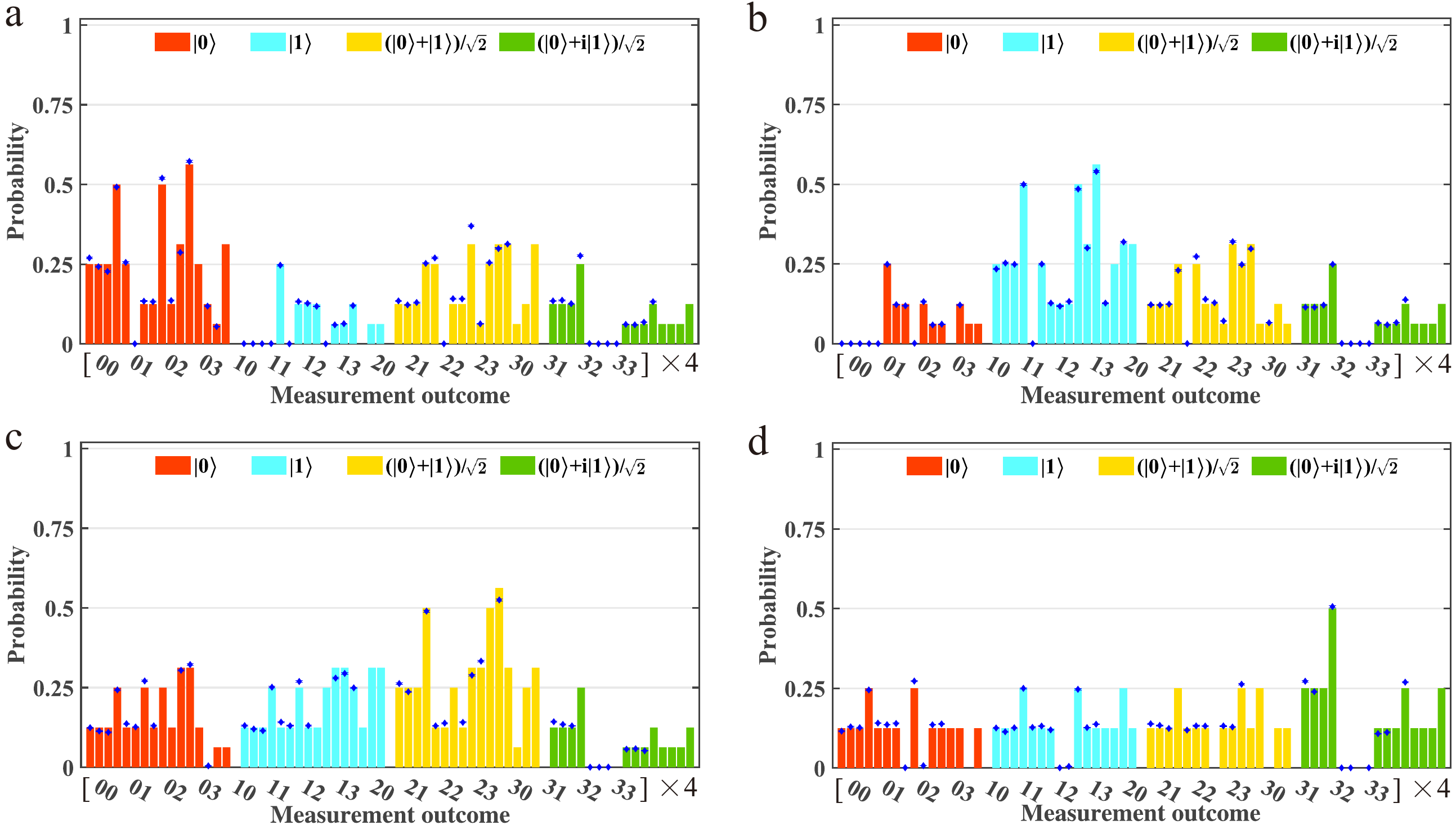}
    \caption{\emph{Experimental data for the optimal witness  (settings $|0\>\<+i|$,  $|1\>\<+i|$,  $|+\>\<+i|$, and $|+i\>\<+i|$).}    
      The figure shows the outcome probabilities of different measurements on the control and target qubits, in the setting where  the target qubit is measured on the state $|+i\>$ and re-prepared in the state  
      $|0\>$ ({\bf a}), $|1\>$ ({\bf b}), $|+\>$ ({\bf c}),  and 
      $|+i\>$ ({\bf d}).   The data representation in this figure follows the same convention as in Fig. \ref{fig:smfig1}.}\label{fig:smfig4}
\end{figure*}

\begin{figure*}[ht!]
    \centering
    \includegraphics[width=0.9\linewidth]{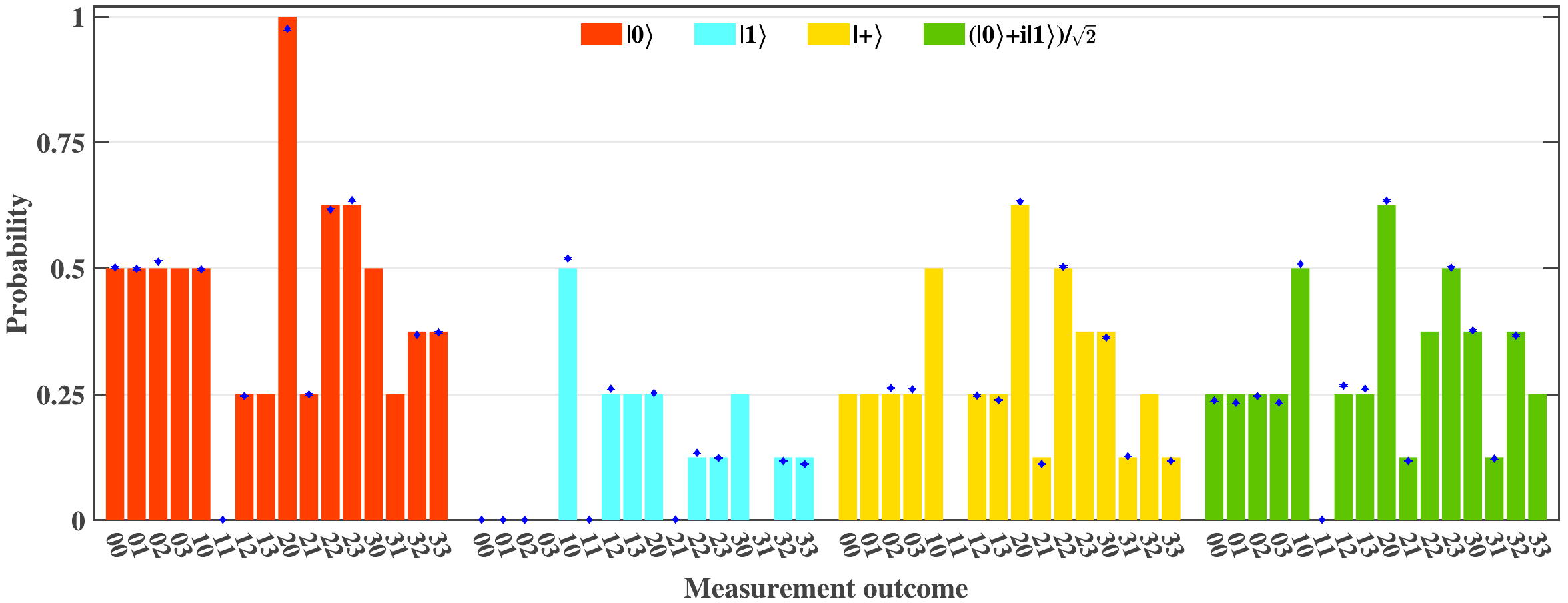}
    \caption{\emph{Probabilities in the decomposition of the simplified witness $W'$.} The dots represents the experimentally observed probabilities. The corresponding theoretical predictions shown in colored bars (red, cyan, yellow, and green bars represent results when the target state is reprepared to $|0\>, |1\>, |+\>, \frac{1}{\sqrt{2}}(|0\>+i|1\>)$ respectively). The $x$ axis enumerates the outcomes of measurements on the control qubit and of the switched measurement on the target qubit, where `0', `1', `2', `3' are used to represent the states  $|0\>, |1\>, |+\>,(|0\>+i|1\>)/\sqrt{2}$ respectively.
    }
    \label{fig:witnessresult2}
\end{figure*}

\section{Details on the  source and   measurement}\label{app:SPS&UG}

All the experiments reported in this paper  used a heralded single photon source based on  a spontaneous parametric down-conversion (SPDC) process on a type-II cut ppKTP crystal.  The crystal was pumped by focusing a 2.5~mW diode laser centered at 404~nm on it using a convex lens (focal length is 12.5 cm). Setting the polarization of the pump laser to be horizontal, we generated pairs of correlated photons centered at 808~nm in a polarization state $|H\>|V\>$, which were then separated by a PBS. The pump laser was blocked with long pass and narrow band pass filters. After this, the photon pairs were coupled into single-mode fibers and detected with single photon detectors (photon counting module from PerkinElmer). The idler photon was used as a herald and the signal photon was sent to our experiment. When setting the coincidence window to be 1 ns, the observed coincidence rate of the photon source was about 20000 pairs per second, the counting rate of each detector was about 60000, and thus the coincidence efficiency was 0.33. The coincidence rate was attenuated to 1850 pairs per second after the signal photon passed through the whole apparatus.

An important factor in our experiments is to guarantee a high interference visibility in the Mach-Zehnder interferometer, which in the case of the game is directly related to the success probability.   The coherence length of our photon source was over 1000~$\mu m$. The length difference between the two interference paths was ensured to be within the coherence length by using a trombone-arm delay line composed of a translation stage with a precision of 10~$\mu m$. The phase between the two interference paths was stabilized by using a piezoelectric transducer (not shown in Fig.~2 of the main text). The interference visibility was measured to be $0.9921\pm 0.0035$ in our experiment (see the end of Section \ref{app:expgame} for more details).  

It is also worth mentioning that our setup guarantees a high single-photon purity, which can be estimated from the expression  $\sqrt{1-g^{(2)}(0)}$, where  $g^{(2)}(0)$ is the heralded idler-idler self-correlation.    In the end of Section \ref{app:expgame},  we briefly discuss how   $g^{(2)}(0)$ can be estimated from our experimental data (including coincidence rates between detectors, coincidence efficiency, and coincidence window). The estimation  yields the value $g^{(2)}(0)\approx 0.0003$, corresponding to a single photon purity of 0.9998.

To witness single-device  input-output indefiniteness, we  projected the control qubit (encoded in the spatial modes of the single photon) onto four states $|0\>$, $|1\>$, $|+\>$, and $|+i\>:  = (|0\>+i|1\>)/\sqrt{2}$. This is done by using two liquid crystal variable retarder (LC1 and LC2 in Fig. 2 of the main text) and a beam splitter (BS2). We use BS2 to choose which spatial mode is measured. When BS2 is removed, measurements onto $|0\>$ and $|1\>$ can be achieved on Port 0 and Port 1 respectively. To achieve measurements onto $|+\>$ and $|+i\>$, on the other hand, we use BS2 to recombine the two spatial modes and add a relative phase of $0$ or $\pi$ between these two spatial modes. The relative phase is achieved by using the two LCs. Specifically, we align the optical axes of LC1 and LC2 with the horizontal and vertical directions respectively and apply voltages on the two LCs to introduce a phase on the horizontally and vertically polarized photons. When the phases added on the horizontally and vertically polarized photons are the same, we can realize adding a phase on the control qubit while not influencing the target qubit.

The above setup can be used to estimate any witness of input-output indefiniteness. This is done by expanding the witness on a set of state preparation and measurement settings as described in Sec.IV. For each setting, we collected the data for 15 seconds and the outcome probabilities were then estimated. The probabilities used in the calculation of the optimal witness $W^{\rm opt}$ are shown in Figs. (\ref{fig:smfig1}-\ref{fig:smfig4}). These probabilities correspond to different settings, where different measure-and-reprepare devices are inserted in our setup, and different events are registered.   For a  measure-and-reprepare device that measures on the basis $\{|a_0\>,  |a_1\> \}$ and reprepares in the basis $\{  |b_0\>, |b_1\>\}$, we label the possible events by the operators $|b_i\>\<a_i|$,  $i\in  \{0,1\}$.  

Similarly, the probabilities used in the experimental evaluation of the simplified witness $W'$ are shown in Fig. (\ref{fig:witnessresult2}).

\section{Witnesses for the QTF}

Here we use the SDP  Eq. (\ref{eq:robustness_sdp}) to calculate the optimal witness of input-output indefiniteness for the quantum time flip supermap  $\map F$, with the control qubit initialized in  the state  $|+\> = (|0\> + |1\>)/\sqrt{2}$. 

Consider the case where the target system is a qubit.  The input (output) qubit of a bistochastic channel $\map C$ will be denoted by $A_{\rm I}$ ($A_{\rm O}$), while the input (output) qubits of $\map F (\map C)$ will be denoted by $B_{\rm I} =  B_{\rm it} B_{\rm ic}$  ($B_{\rm O} =  B_{\rm ot} B_{\rm oc}$), where $B_{\rm it}$  ($B_{\rm ot}$) is the input (output) target qubit and  $B_{\rm ic}$ ($B_{\rm oc}$) is the input (output) control qubit. 
 The Choi operator of QTF  is given by~\cite{chiribella2022quantum}
\begin{equation}
    \label{eq:choi_qtf}
    \op{Choi}(\map F) = |V\>\< V| \, ,
\end{equation}
with
\begin{equation}
    \begin{split}
        |V\> = &\, |I\kk_{A_{\rm I}B_{\rm it}} \otimes |I\kk_{A_{\rm O}B_{\rm ot}} \otimes |0\>_{B_{\rm ic}} \otimes |0\>_{B_{\rm oc}} \\
            &+ |I\kk_{A_{\rm O}B_{\rm it}} \otimes |I\kk_{A_{\rm I}B_{\rm ot}} \otimes |1\>_{B_{\rm ic}} \otimes |1\>_{B_{\rm oc}} \, .
    \end{split}
\end{equation}
It follows that the  Choi operator including the state preparation of the control qubit in the state $|+\>$ is
\begin{equation}
    S_{{\rm QTF},  |+\>\<+|} = |V^+\>\<V^+| \, ,
\end{equation}
with
\begin{equation}
    \begin{split}
        |V^+\> = &\, \frac 1 {\sqrt 2} |I\kk_{A_{\rm I}B_{\rm it}} \otimes |I\kk_{A_{\rm O}B_{\rm ot}} \otimes |0\>_{B_{\rm oc}} \\
            &+ \frac 1 {\sqrt 2} |I\kk_{A_{\rm O}B_{\rm it}} \otimes |I\kk_{A_{\rm I}B_{\rm ot}} \otimes |1\>_{B_{\rm oc}} \, .
    \end{split}
\end{equation}
Solving the SDP  (\ref{eq:robustness_sdp}) for $S = S_{{\rm QTF},  |+\>\<+|}$, we then obtain the robustness  $r(S_{{\rm QTF},  |+\>\<+|}) \approx 0.4007$ and a matrix representation of the optimal witness $W^{\rm opt}$.  To measure the witness $W^{\rm opt}$ in the experiment, we decompose it into a collection of linearly independent operations, including state preparations for $B_{\rm it}$,  measurements on  $A_{\rm I}$, state repreparations of  $A_{\rm O}$, and measurements on $B_{\rm ot}$ and $B_{\rm oc}$. To be bidirectional, the measure-and-reprepare operations are required to be bistochastic instruments \cite{chiribella2021symmetries}, i.e. the operators $\{M_j\}$ of such an instrument satisfy the condition that $\sum_j M_j$ is a bistochastic channel. In the case of QTF, we realized bistochastic instruments by measuring the system $A_{\rm I}$ in some orthonormal basis $\{|v_0\>, |v_1\>\}$, and then repreparing states from another orthonormal basis $\{ |w_0\>, |w_1\> \}$, with the state repreparation depending on the measurement outcome.  Explicitly, the decomposition is $W^{\rm opt} = \sum_{a,b,c,d,e} \alpha_{a,b,c,d,e}   \,  W_{abcde}$, where $\alpha_{a,b,c,d,e}$ are real coefficients (obtained from the solution of the SDP), 
\begin{multline}
    W_{abcde} =  |\overline a\>\<\overline a|_{B_{\rm it}} \otimes | b\>\< b|_{A_{\rm I}} \otimes |\overline c\>\<\overline c|_{A_{\rm O}} \\\otimes | d\>\< d|_{B_{\rm ot}} \otimes |e\>\< e|_{B_{\rm oc}} \, ,
\end{multline}
and the vectors $|a\>$, $|b\>$, $|c\>$, $|d\>$ and $|e\>$ are chosen from the set
\begin{equation}
    \label{eq:operation_basis}
    |0\>, |1\>, \frac{|0\>+|1\>}{\sqrt 2}, \frac{|0\>+i|1\>}{\sqrt 2} \, .
\end{equation}
The number of terms that contribute to the decomposition of $W^{\rm opt}$ is 794. 

The robustness is then given by the expression 
\begin{equation}
    \begin{split}
    r(S_{{\rm QTF}, |+\>\<+|}) &= -\Tr(W^{\rm opt}S_{{\rm QTF}, |+\>\<+|}) \\
    &= -\sum_{a,b,c,d,e}  \alpha_{a,b,c,d,e} \,  p(a,b,c,d,e)\, ,
    \end{split}
\end{equation}
where $p(a,b,c,d,e) =  \Tr[ W_{abcde}  \, S_{{\rm QTF},  |+\>\<+|} ]$ is the probability of the event labeled by $(a,b,c,d,e)$. In the experiment, we estimated the probabilities $p(a,b,c,d,e)$, and inserted the estimates into the expression of the robustness, obtaining the experimental value $0.345 \pm 0.005$.

We also consider witnesses of input-output indefiniteness that require fewer measurement settings. To this purpose, we restrict the optimization of the robustness to a subset of witnesses the can be estimated by measuring only $A_{\rm I}$, $A_{\rm O}$, $B_{\rm oc}$, fixing the state of $B_{\rm it}$ to  $|0\>$ and tracing out $B_{\rm ot}$. These witnesses can be computed by adding the following constraint to Eq. (\ref{eq:robustness_sdp}):
\begin{equation}
    \label{eq:reduce_witness}
    W = |0\>\<0|_{B_{\rm it}} \otimes I_{B_{\rm ot}} \otimes W^{\rm reduce} \, ,
\end{equation}
where $W^{\rm reduce}$ is a Hermitian matrix on $A_{\rm I} \otimes A_{\rm O} \otimes B_{\rm oc}$. The maximal robustness under this constraint is given by $r'  (S_{{\rm QTF},  |+\>\<+|})  \approx  0.1716$. The witness that achieves  maximal robustness, denoted by $W'$, can be decomposed as  $W' = \sum_{b,c,e} \beta_{b,c,e} \,  W'_{bce}$, where $\beta_{b,c,e}$ are real coefficients,   $W'_{bce} =  |0\>\<0|_{B_{\rm it}} \otimes I_{B_{\rm ot}}  \otimes |\overline b\>\<\overline b|_{A_{\rm I}} \otimes |c\>\<c|_{A_{\rm O}} \otimes |\overline e\>\<\overline e|_{B_{\rm oc}}$,  and the vectors $|b\>$, $|c\>$, and $|e\>$ are chosen from the set (\ref{eq:operation_basis}).  Now, the number of terms that contribute to the decomposition is only 48.   The robustness is  $r'  (S_{{\rm QTF},  |+\>\<+|}) =  \sum_{b,c,e}  \,\beta_{b,c,e} \,  p(b,c,e)$,  with $p(b,c,e)   =  \Tr[ W_{bce}' \,  S_{{\rm QTF},  |+\>\<+|}]$. 
 By experimentally estimating the probabilities $p(b,c,e)$, we then obtain the experimental value   $0.140 \pm 0.004$. 
 
We show that 3 is the minimum number of qubits to be measured for witnesses of input-output indefiniteness of QTF. Thus the complexity of witnesses of QTF can not be further reduced.

 \begin{lemma}
    \label{theo:singleslot}
    Supermaps transforming bistochastic channels to unit probability has no indefiniteness of input-output direction.
\end{lemma}
\Proof Let $S$ be the Choi operator of a supermap transforming every bistochastic channel (from system $A_{\rm I}$ to system $A_{\rm O}$) to unit probability. Then $S$ is a postive operator on $\spc H_{A_{\rm I}} \otimes \spc H_{A_{\rm O}}$ satisfying $S = {}_{[A_{\rm I}]}S + {}_{[A_{\rm O}]}S - {}_{[A_{\rm I}A_{\rm O}]}S$. Consider the spectral decomposition
\begin{align}
    {}_{[A_{\rm O}]}S &= \sum_i \lambda_i |u_i\>\<u_i| \otimes I_{A_{\rm O}} \, , \\
    {}_{[A_{\rm I}]}S &= I_{A_{\rm I}} \otimes \sum_i \mu_j |v_j\>\<v_j| \, .
\end{align}
Note that ${}_{[A_{\rm I}]}S$ and ${}_{[A_{\rm O}]}S$ are simultaneously diagonalizable. Let ${}_{[A_{\rm I}A_{\rm O}]}S = m\, I_{A_{\rm I}} \otimes I_{A_{\rm O}}$. Positivity of $S$ implies that, for any $i,j$, the sum of eigenvalues satisfies $\lambda_i + \mu_j \geq m$ and thus the minimum satisfies $\lambda_{\min} + \mu_{\min} \geq m$. Now we can decompose $S$ into the sum of two positive operators: $S = S_f + S_b$ where
\begin{equation}
    S_f = {}_{[A_{\rm O}]}S - \lambda_{\min} I_{A_{\rm I}} \otimes I_{A_{\rm O}}
\end{equation}
and
\begin{equation}
    S_b = {}_{[A_{\rm I}]}S - (m - \lambda_{\min}) I_{A_{\rm I}} \otimes I_{A_{\rm O}} \, .
\end{equation}
It holds that $S_f = {}_{[A_{\rm O}]}S_f$ and $S_b = {}_{[A_{\rm I}]}S_b$. Therefore the supermap $S$ is a random mixture of a forward and backward supermaps (corresponding to the operators $S_f$ and $S_b$, respectively).
\qed

\begin{theo}
    Any witness of QTF has to include variations on both system $A_{\rm I}$ and system $A_{\rm O}$, as well as one of the control systems: $B_{\rm ic}$ and $B_{\rm oc}$.
\end{theo}
\Proof
If there is no variation on the composite system of $A_{\rm I}$ and $A_{\rm O}$, i.e. a fixed bistochastic channel $\map C_0$ is applied on system $A_{\rm I}$ and $A_{\rm O}$, then the reduced process on the remaining systems is a simple channel from $B_{\rm it}B_{\rm ic}$ to $B_{\rm ot}B_{\rm oc}$, which is irrelevant to the input-output direction of $A_{\rm I}$ and $A_{\rm O}$.
If there is no variation on system $A_{\rm O}$ in a witness $W$, i.e. $W = \Tr_{A_{\rm O}} W \otimes \frac{I_{A_{\rm O}}}{d_{A_{\rm O}}}$. Then the expectation of $W$ on $S_{\rm QTF}$ satisfies
\begin{equation}
    \Tr(WS_{\rm QTF}) = \Tr \left[ W \left(\frac 1 {d_{A_{\rm O}}} \Tr_{A_{\rm O}}S_{\rm QTF} \otimes I_{A_{\rm O}} \right) \right] \, .
\end{equation}
The reduced operator $\frac 1 {d_{A_{\rm O}}}  {\Tr_{A_{\rm O}}S_{\rm QTF}}$ turns out to be the the Choi operator of a channel $\map C$ from $B_{\rm it}B_{\rm ic}$ to $A_{\rm I}B_{\rm ot}B_{\rm oc}$ since $\frac 1 {d_{A_{\rm O}}} \Tr_{A_{\rm I}A_{\rm O}B_{\rm ot}B_{\rm oc}}S_{\rm QTF} = I_{B_{\rm it}} \otimes I_{B_{\rm ic}}$. 
Hence the operator $\frac 1 {d_{A_{\rm O}}}\Tr_{A_{\rm O}}S_{\rm QTF} \otimes I_{A_{\rm O}}$ corresponds to a forward supermap consisting of the channel $\map C$ and a discarding operation on system $A_{\rm O}$. It follows that the expectation $\Tr(WS_{\rm QTF})$ must be non-negative. Similarly, if there is there is no variation on system $A_{\rm I}$, the expectation of witnesses must be non-negative. Thus, variations on both system $A_{\rm I}$ and system $A_{\rm O}$ is necessary for certification of input-output indefiniteness of QTF.

In the case that variations occur only on systems $A_{\rm I}$ and $A_{\rm O}$, according to Lemma \ref{theo:singleslot}, the reduced operator of any supermap corresponds to a supermap transforming bistochastic channels to unit probability and thus has no indefiniteness of input-output direction. Hence variations restricted to systems $A_{\rm I}$ and $A_{\rm O}$ are not sufficient.

Now we show that a variation on one of the control systems of QTF is necessary. If the control qubit $B_{\rm ic}$ of QTF is initialized in a fixed state $\sigma$ (or the target $B_{\rm it}$ and control system $B_{\rm ic}$ are initialized in a joint state $\rho$), and the control qubit $B_{\rm oc}$ is traced out in the end, then QTF becomes a random mixture of a forward supermap and backward supermap. To see this, we compute the Choi operator of the supermap after fixing the corresponding deterministic operations on systems $B_{\rm ic}$ ($B_{\rm it}B_{\rm ic}$) and system $B_{\rm oc}$:
\begin{align}
    &\Tr_{B_{\rm ic}B_{\rm oc}} (S_{\rm QTF} \, \sigma_{B_{\rm ic}}^T) \nonumber \\
    &= \, \<0|\sigma|0\> \cdot |I\kk\bb I|_{A_{\rm I}B_{\rm it}} \otimes |I\kk\bb I|_{A_{\rm O}B_{\rm ot}} \nonumber \\
    &\quad + \<1|\sigma|1\> \cdot |I\kk\bb I|_{A_{\rm O}B_{\rm it}} \otimes |I\kk\bb I|_{A_{\rm I}B_{\rm ot}} \\
    &\Tr_{B_{\rm it}B_{\rm ic}B_{\rm oc}} (S_{\rm QTF} \, \rho_{B_{\rm it}B_{\rm ic}}^T) \nonumber \\
    &= \, P_0 \otimes |I\kk\bb I|_{A_{\rm O}B_{\rm ot}}  + P_1 \otimes |I\kk\bb I|_{A_{\rm I}B_{\rm ot}} \, .
\end{align}
where
\begin{equation}
    P_0 = \Tr_{B_{\rm it}B_{\rm ic}} [(|I\kk\bb I|_{A_{\rm I}B_{\rm it}} \otimes |0\>\<0|_{B_{\rm ic}}) \rho_{B_{\rm it}B_{\rm ic}}^T]
\end{equation}
and
\begin{equation}
    P_1 = \Tr_{B_{\rm it}B_{\rm ic}} [(|I\kk\bb I|_{A_{\rm O}B_{\rm it}} \otimes |1\>\<1|_{B_{\rm ic}}) \rho_{B_{\rm it}B_{\rm ic}}^T]
\end{equation}
are positive operators.

In conclusion, any witness of QTF has to include variations on both system $A_{\rm I}$ and system $A_{\rm O}$, as well as one of the control systems: $B_{\rm ic}$ and $B_{\rm oc}$.
\qed



\section{Bipartite witnesses of input-output indefiniteness}\label{app:game_witness}

Here we introduce the notion of bipartite witness of input-output indefiniteness,  distinguishing between two levels of strength of this notion and showing that the quantum game discussed in the main text is an example of the weaker kind, while the single-device witnesses introduced earlier in our paper provide examples of the stronger kind.

\subsection{Strong vs weak witnesses}

Consider a setup that uses a pair of bidirectional devices (mathematically represented by their forward processes $\map A$ and $\map B$, respectively) to generate a new device (mathematically represented by a quantum channel;  $\map C$). The setup can be represented by a Choi operator, acting on the Hilbert spaces of systems $A_{\rm I}, A_{\rm O}$ (input and output of device $\map A$), $B_{\rm I}, B_{\rm O}$ (input and output of device $\map B$), and $C_{\rm I}, C_{\rm O}$ (input and output of device $\map C$).    Mathematically, the setup is a bilinear supermap, sending pairs of bistochastic channels $(\map A, \map B)$ into (generally non-bistochastic) channels $\map C$.  These supermaps can be naturally extended to supermaps transforming no-signaling bipartite bistochastic channels, of the form $\map N  =  \sum_i \, x_i \map A_i\otimes \map B_i $ where $\map A_i$ and $\map B_i$ are bistochastic and $x_i$ are real coefficients,  into channels $\map C$.      The set of Choi operators of these supermaps has been characterized in Ref. \cite{chiribella2022quantum}.

Now, consider the subset of bipartite supermaps that use both devices $\map A$ and $\map B$ in the forward direction,  meaning that the action of these supermaps is well-defined even if $\map A$ and $\map B$ are ordinary (non-bistochastic) channels.   These supermaps coincide with the supermaps defined in Refs. \cite{oreshkov2012quantum,chiribella2013quantum}, where the input-output direction is fixed and the relative order of the devices $\map A$ and $\map B$ can be indefinite. 
The Choi operators of these supermaps satisfy the constraints \cite{oreshkov2012quantum} \begin{align}
  \nonumber  S \geq 0 \, , \\
 \nonumber   \Tr_{A_IA_OB_IB_O}S = d_A d_B I_{C_I} \, ,\\
  \nonumber   {}_{[A_IA_OC_O]}S = {}_{[A_IA_OB_OC_O]}S \, ,\\
  \nonumber  {}_{[B_IB_OC_O]}S = {}_{[A_OB_IB_OC_O]}S \, ,\\
    {}_{[C_O]}S = {}_{[A_OC_O]}S - {}_{[B_OC_O]}S + {}_{[A_OB_OC_O]}S \, .\label{fwfw}
\end{align}  
We denote the set of Choi operators satisfying these constraints as $\set{S}^{\rm fw, fw}_{\rm definite}  $,  meaning that they are Choi operators of supermaps that use both devices $\map A$ and $\map B$ in the forward direction.  

Similarly, we define the set $\set{S}^{\rm bw,\rm bw}_{\rm definite}$ of (Choi operators of) supermaps that use both devices in the backward direction.  Mathematically, the set $\set{S}_{\rm bw,\rm bw}$ is characterized by the constraints
\begin{align}
  \nonumber    S \geq 0 \, , \\
   \nonumber   \Tr_{A_IA_OB_IB_O}S = d_A d_B I_{C_I} \, ,\\
   \nonumber   {}_{[A_IA_OC_O]}S = {}_{[A_IA_OB_IC_O]}S \, ,\\
  \nonumber    {}_{[B_IB_OC_O]}S = {}_{[A_IB_IB_OC_O]}S \, ,\\
      {}_{[C_O]}S = {}_{[A_IC_O]}S + {}_{[B_IC_O]}S - {}_{[A_IB_IC_O]}S \, ,
  \end{align}  
which can be obtained from the constraints (\ref{fwfw}) by exchanging the roles of the input and output systems.  

The sets $\set S^{\rm fw, fw}_{\rm definite}$ and $\set S^{\rm bw, bw}_{\rm definite}$ are compatible with a global input-output direction, defined jointly for both devices $\map A$ and $\map B$.   These sets are especially important in scenarios where the input-output direction coincides with the arrow of time.  In this case, the set $\set S^{\rm fw, fw}_{\rm definite}$ represents the largest set of operations accessible to an agent that operates in the forward time direction, while the set $\set S^{\rm bw, bw}_{\rm definite}$ represents the largest set of operations accessible to a hypothetical agent that operates in the backward time direction.

In general, however, one can also consider setups that use device $\map A$ in the forward direction and device $\map B$ in the backward direction, or {\em vice-versa}. The corresponding sets of Choi operators, denoted by  $\set S^{\rm fw, bw}_{\rm definite}$ and  $\set S^{\rm bw, fw}_{\rm definite}$, are characterized by the constraints    
\begin{align}
  \nonumber  S \geq 0 \, , \\
   \nonumber   \Tr_{A_IA_OB_IB_O}S = d_A d_B I_{C_I} \, ,\\
 \nonumber   {}_{[A_IA_OC_O]}S = {}_{[A_IA_OB_IC_O]}S \, ,\\
  \nonumber  {}_{[B_IB_OC_O]}S = {}_{[A_OB_IB_OC_O]}S \, ,\\
    {}_{[C_O]}S = {}_{[A_OC_O]}S + {}_{[B_IC_O]}S - {}_{[A_OB_IC_O]}S \, 
\end{align}  
and \begin{align}
  \nonumber  S \geq 0 \, , \\
   \nonumber   \Tr_{A_IA_OB_IB_O}S = d_A d_B I_{C_I} \, ,\\
 \nonumber   {}_{[A_IA_OC_O]}S = {}_{[A_IA_OB_OC_O]}S \, ,\\
  \nonumber  {}_{[B_IB_OC_O]}S = {}_{[A_IB_IB_OC_O]}S \, ,\\
    {}_{[C_O]}S = {}_{[A_IC_O]}S + {}_{[B_OC_O]}S - {}_{[A_IB_OC_O]}S \, ,
\end{align}  
respectively.

Now, consider the setups that use both devices in a definite input-output direction, with the same input-output direction for both devices.   The corresponding set of Choi operators, denoted by $\set{S}_{\rm definite}^{\rm same}$,  consists of all Choi operators of the form 
\begin{align}
S_{\rm same}  =  p  \,  S_{\rm fwd, fwd}   +  (1-p)\,  S_{\rm bwd, bwd}\, ,
\end{align}  
where $p\in  [0,1]$ is a probability, and  the operators $S_{\rm fwd, fwd}, \,  S_{\rm bwd, bwd}$   belong to the sets   $\set{S}^{\rm fwd, fwd}_{\rm definite},  \set{S}^{\rm bwd, bwd}_{\rm definite}$, respectively.

Likewise, we can consider the setups that use the two devices in a definite input-output direction, but with opposite directions for the two devices.  The corresponding set of Choi operators, denoted by $\set{S}_{\rm definite}^{\rm same}$,  consists of all Choi operators of the form 
\begin{align}
S_{\rm opposite}   =  p  \,  S_{\rm fwd, bwd}   +  (1-p)\,  S_{\rm bwd, fwd}\, ,
\end{align}  
where $p\in  [0,1]$ is a probability, and  the operators $S_{\rm fwd, bwd}, \,  S_{\rm bwd, fwd}$   belong to the sets   $\set{S}^{\rm fwd, bwd}_{\rm definite},  \set{S}^{\rm bwd, fwd}_{\rm definite}$, respectively.   

Finally, we can consider the setups that use both devices in a definite input-output direction, without any restriction on how the input-output direction of the first device is related to the input-output direction of the second device.  The Choi operators of these setups are of the form
\begin{align}\label{everydefinite}
S_{\rm definite}   =  p  \,  S_{\rm same}   +  (1-p)\,  S_{\rm  opposite}\, ,
\end{align}  
where $p\in  [0,1]$ is a probability, and  the operators $S_{\rm same}, \,  S_{\rm opposite}$   belong to the sets   $\set{S}^{\rm same}_{\rm definite},  \set{S}^{\rm opposite}_{\rm definite}$, respectively.   
The set of Choi operators of the form  (\ref{everydefinite}) will be denoted by $\set S_{\rm definite}$.

We are now ready to define three different notions of witnesses of bipartite input-output indefiniteness.   

\begin{defi}\label{defi:weak1}
An operator $W$  acting on the composite system $A_IA_OB_IB_OC_IC_O$ is a {\em weak witness of bipartite input-output indefiniteness} if  $\Tr[W  S] \ge 0$ for every $S\in \set{S}^{\rm same}_{\rm definite}$ and $\Tr[W  S]  <0$ for some $S  \in \set S$. 
\end{defi}

\begin{defi}\label{defi:weak2}
An operator $W$  acting on the composite system $A_IA_OB_IB_OC_IC_O$ is a {\em conjugate weak witness of bipartite input-output indefiniteness} if  $\Tr[W  S] \ge 0$ for every $S\in \set{S}^{\rm opposite}_{\rm definite}$ and $\Tr[W  S]  <0$ for some $S  \in \set S$. 
\end{defi}

\begin{defi}\label{defi:strong}
An operator $W$  acting on the composite system $A_IA_OB_IB_OC_IC_O$ is a {\em strong witness of bipartite input-output indefiniteness} if  $\Tr[W  S] \ge 0$ for every $S\in \set{S}_{\rm definite}$ and $\Tr[W  S]  <0$ for some $S  \in \set S$.   
\end{defi}

The above definitions imply that an operator $W$ is a strong witness if and only if $W$ is both a weak witness and a conjugate weak witness.

\subsection{The game as a weak witness}

We now show that the game described in the main text is a weak witness, but not a strong one.  

First, let us cast the game in the form of a witness.  The possible strategies are described by bipartite supermaps which have two slots (corresponding to systems $A_{\rm I}A_{\rm O}$ and $B_{\rm I}B_{\rm O}$) and one qubit output $C_{\rm O}$  (compared to the general framework in the previous subsection, here we are taking the input system $C_{\rm I}$ to be trivial). The strategy is carried out by placing the two gates $U$ and $V$ in the slots and then measuring the output qubit $C_{\rm O}$ in Fourier basis $\{ |\pm\> \}$ which gives the outcome that $(U, V)$ belongs to
\begin{equation}
    \set G_+ = \{ (U, V) \mid UV^T = U^TV \} \, ,
\end{equation}
or
\begin{equation}
    \set G_- = \{ (U, V) \mid UV^T = -U^TV \} \, .
\end{equation}
Let $S$ be the Choi operator of the strategy. If the gates $(U, V) \in \set G_+$, then the probability of winning is
\begin{equation}
    \Tr((\op{Choi}(U) \otimes \op{Choi}(V) \otimes |+\>\<+|)^T S) \, .
\end{equation}
If the gates $(U, V) \in \set G_-$, then the probability of winning is
\begin{equation}
    \Tr((\op{Choi}(U) \otimes \op{Choi}(V) \otimes |-\>\<-|)^T S) \, .
\end{equation}
In the general case, suppose that $\mu(U, V)$ is a probability measure on $\set G_+ \cup \set G_-$. We define the operators $M_+$ and $M_-$ to be
\begin{equation}
    M_\pm := \int_{\set G_{\pm}} (\op{Choi}(U) \otimes \op{Choi}(V) \otimes |\pm\>\<\pm|)^T \d \mu(U, V) \, .
\end{equation}
Then the probability of winning is
\begin{multline}
    p_{\rm succ} = \\
    \int_{\set G_+} \Tr((\op{Choi}(U) \otimes \op{Choi}(V) \otimes |+\>\<+|)^T S)
    \cdot \d \mu(U, V) \\
    + \int_{\set G_-} \Tr((\op{Choi}(U) \otimes \op{Choi}(V) \otimes |-\>\<-|)^T S)
    \cdot \d \mu(U, V) \\
    = \Tr((M_+ + M_-)S) \, .
\end{multline}

 Ref. \cite{chiribella2022quantum}  provided examples of probability distributions $\d \mu  (U,V)$ with the property that the probability of winning is strictly smaller than 1 for every strategy that uses both gates in the forward direction, or both gates in the backward direction.    
 For all those probability distributions,  the operator \begin{equation}
    W_{\rm game} := \frac I {d^2} - \frac {M_+ + M_-} {p_{\max}} \, ,
\end{equation}
is a weak witness.

The witness $ W_{\rm game}$ is not a strong witness,  because there exist quantum strategies that achieve a unit probability of success while using each of the gates  $U$ and $V$ in definite (opposite) input-output directions.

For example, a perfect winning strategy uses the gate $U$ in the forward direction and the gate $V$ in the backward direction, corresponding to the gate $V^T$.    The strategy is to insert the gates $U$ and $V^T$ into the  quantum SWITCH supermap \cite{chiribella2013quantum}, which turns them into the controlled unitary gate 
\begin{align}
S(U,V^T)   =   UV^T  \otimes |0\>\<0|  +  V^T  U \otimes |1\>\<1|  \,.     
\end{align}
Now, if the control qubit of the quantum SWITCH is initialized in the  state $|+\> = (|0\>+|1\>)/\sqrt{2}$ and the target system is maximally entangled with an auxiliary system $R$,   the gate $S(U, V^T)$ produces the output state
\begin{equation}
\frac{|UV^T + V^TU\kk}{2\sqrt d} \otimes |+\> + \frac{|UV^T - V^TU\kk}{2\sqrt d} \otimes |-\>, 
\end{equation}
where  we used the double-ket notation 
\begin{equation}
    |M\kk =    (M\otimes I)\,  |I\kk \, .
\end{equation}
We now show that a projective measurement on the above states can determine with certainty whether the gates $(U, V)$ belong to the $\set G_+$ or to the $\set G_-$.   

Suppose that $(U, V)$ belongs to $ \set G_+$. In this case,  we obtain the relations
\begin{align}
UV^T +   V^TU   =   UV^T +    V  U^T    \in  \spc H_+  \,,
\end{align}
and \begin{align}
UV^T -   V^TU   =   UV^T -    V  U^T    \in  \spc H_-  \,,
\end{align}
where $\spc H_\pm$ are the symmetric and anti-symmetric subspaces of $\spc H\otimes \spc H$.  

Instead, if $(U, V)$ belongs to $ \set G_-$, we have 
\begin{align}
UV^T +   V^TU   =   UV^T -    V  U^T    \in  \spc H_-  \,,
\end{align}
and \begin{align}
UV^T -   V^TU   =   UV^T +    V  U^T    \in  \spc H_+  \,,
\end{align}

Hence, it is possible to determine whether the pair  $(U, V)$ belongs to $ \set G_+$ or to the set $ \set G_-$ by performing the binary projective measurement with projectors 
\begin{align}
Q_+   =   P_+  \otimes |+\>\<+|   +   P_-  \otimes |-\>\<-|
\end{align}
and 
\begin{align}
Q_-   = P_-  \otimes |+\>\<+|   +   P_+ \otimes |-\>\<-| \, .
\end{align}
The  (Choi operator of the) above winning strategy belongs to the set $\set{S}_{\rm definite}^{\rm opposite}$.

\begin{figure*}[ht!]
    \centering
    \includegraphics[width=0.4\linewidth]{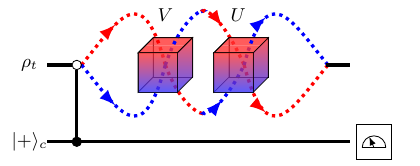}
    \caption{\emph{Theoretical scheme for the game.} A target system traverses two devices V and U in two different directions dependent on the state of a control qubit, which is prepared in the state $|+\>$. To be explicit, the target undergoes $U^TV$ if the control is in $|0\>$ (red path) and undergoes $UV^T$ if the control is in $|1\>$ (blue path).
    }
    \label{fig:gamescheme}
\end{figure*}
\begin{figure*}[ht!]
    \centering
    \includegraphics[width=0.7\linewidth]{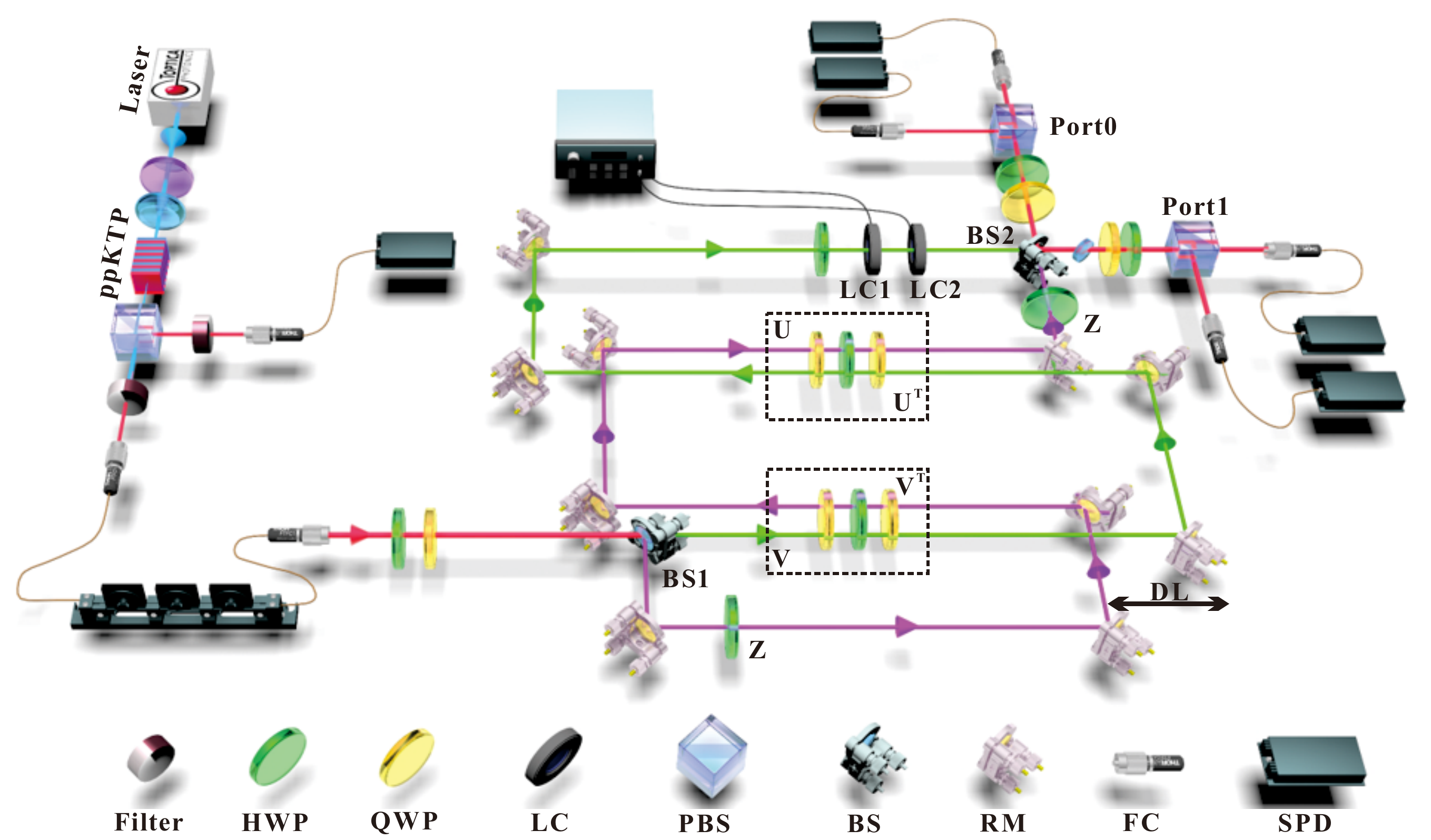}
    \caption{\emph{Experimental setup for the game.} The source and measurements are the same as in the experiment on the witness. The only difference is that instead of the measure-and-reprepare process, we put two unitary gates $U$ and $V$ in the setup.  The abbreviations used in the figure are:  HWP, half-wave plate; QWP, quarter-wave plate; PBS, polarizing beam splitter; BS, beam splitter ($T/R=50/50$); RM, reflection mirror; LC, liquid crystal variable retarder; FC, fiber coupler; SPD, single photon detector; DL, trombone-arm delay line.
    }
    \label{fig:gamesetup}
\end{figure*}
 
\subsection{Strong bipartite witnesses from single-device witnesses}  

Single-device witnesses can be easily converted into strong bipartite witnesses.   Intuitively, the idea is that if we can rule out a definite input-output direction for the use of an individual device, then we can also rule out a definite input-output direction for the combined use of two or more devices.  

More formally,   suppose that  $W  \in  L(  \spc H_{A_{\rm I}} \otimes \spc H_{A_{\rm O}} \otimes \spc H_{C_{\rm I}} \otimes \spc H_{C_{\rm O})}$ is a single-device witness for setups that transform an input bistochastic channel $\map A$ (with input $A_{\rm I}$ and output $A_{\rm O}$) into a channel $\map C$ (with input $C_{\rm I}$ and output $C_{\rm O}$), and that $B$ is the Choi operator of a bistochastic channel with input $B_{\rm I}$ and output $B_{\rm O}$. Then, the operator $  W \otimes B $ is a strong witness for bipartite setups taking in input two bidirectional devices (with input-output pairs $(A_{\rm I},  A_{\rm O})$ and $(B_{\rm I},  B_{\rm O})$, respectively) and producing in output a device (with input-output pair $(C_{\rm I},  C_{\rm O})$).  This particular witness corresponds to plugging a fixed bidirectional device into one slot of the bipartite setup, and testing the input-output indefiniteness in the remaining slot. 

Similarly, one can construct strong witnesses of the form $A \otimes W'$, where $A$ is the Choi operator of a bistochastic channel  with input $
A_{\rm I}$ and output $A_{\rm O}$, and $W' \in  L(\spc H_{B_{\rm I}} \otimes \spc H_{B_{\rm O}}  \otimes \spc H_{C_{\rm I}} \otimes \spc H_{ C_{\rm O})}$ is a single-device witness for setups that transform an input bistochastic channel $\map B$ (with input $B_{\rm I}$ and output $B_{\rm O}$) into a channel $\map C$ (with input $C_{\rm I}$ and output $C_{\rm O}$). 

By taking linear combinations of witnesses of the above form, one can easily construct examples of strong bipartite witnesses that take negative values only if both devices are used in an indefinite input-output direction.  It is worth stressing that, of course, not all strong witnesses are of this form.  However, the construction shown in this subsection is conceptually important because it shows that single-device witnesses, like the witnesses measured in our experiments, offer intrinsically stronger certificates of input-output indefiniteness compared to weak bipartite witnesses.

\section{Experimental setup for the quantum game}\label{app:expgame}

The theoretical scheme and the experimental setup for the quantum game are shown in Fig. \ref{fig:gamescheme} and Fig. \ref{fig:gamesetup} respectively. Compared to the single-device witness, the game has a simpler realization. The spatial qubit is initialized in the state $|+\>$ by a 50/50 beamsplitter, while precise initialization of the polarization is not necessary here, since the winning strategy works equally well for every initial state of the target qubit. Moreover, there measure-and-prepare instruments used in the single-device witness are now replaced by unitary gates.   Depending on the state of the spatial qubit, the polarization qubit traverses two optical devices in two alternative directions. The first (second) device rotates the polarization, implementing the unitary gate $U$ ($V$)  in the forward direction, and the gate  $Z U^T Z$ ($Z V^T Z $)  in the backward direction (the Pauli $Z$ in the backward process is then removed by a Pauli $Z$ rotation placed on the appropriate path).   Overall,  the two paths used in the setup result in the two combinations  $U^T V$ and  $U V^T$.  The two paths are then  recombined by a beamsplitter, and  photon detections are performed at its output. The game \cite{chiribella2022quantum}  involves the implementation of several unitary gate pairs $(U, V)$ satisfying either the property $U V^T=  U^T  V$ or the property $U V^T  =  - U^T V$.    In the experiment, we implement the following sets of gates 
\begin{align}
    \label{unitarygate}
    \set S_+ = &\{(I, I), (I, X), (I, Z), (X, I), (X, X), (X, Z), \notag\\
    & (Z, I), (Z, X), (Z, Z), (U_1, V_1), (V_1, U_1), \notag\\
    & (U_2, V_2), (V_2, U_2)\} \, , \\
    \set S_- = &\{(Y, I), (Y, X), (Y, Z), (I, Y), (X, Y), (Z, Y), \notag\\
    & (U_3, V_3), (V_3, U_3)\} \, ,
\end{align}
where $I$ is identity gate and $X$, $Y$, $Z$ are Pauli gates. These gates are implemented using a combination of three waveplates (in a quarter-half-quarter configuration), using the angle settings shown in Table \ref{table:gatesetting}.   

\begin{table*}[ht!]
\setlength{\tabcolsep}{7mm}{
\begin{tabular}{c|c|c|c}
  \hline
    Unitary gate  & QWP1 & HWP &  QWP2 \\
  \hline
  $I$ & $0^\circ$ & $0^\circ$ & $0^\circ$ \\
  \hline
  $X$ & $0^\circ$ & $45^\circ$ & $0^\circ$\\
  \hline
  $Y$ & $90^\circ$ & $45^\circ$ & $0^\circ$\\
  \hline
  $Z$ & $90^\circ$ & $0^\circ$ & $0^\circ$\\
  \hline
  $U1=(X-Y)/\sqrt{2}$ & $45^\circ$ & $67.5^\circ$ & $135^\circ$\\
  \hline
  $V1=(X+Y)/\sqrt{2}$ & $135^\circ$ & $67.5^\circ$ & $45^\circ$\\
  \hline
  $U2=(Z-Y)/\sqrt{2}$ & $0^\circ$ & $22.5^\circ$& $90^\circ$\\
  \hline
  $V2=(Z+Y)/\sqrt{2}$ & $90^\circ$ & $22.5^\circ$ & $0^\circ$\\
  \hline
  $U3=(I-iY)/\sqrt{2}$ & $22.5^\circ$ & $135^\circ$ & $67.5^\circ$\\
  \hline
  $V3=(I+iY)/\sqrt{2}$ & $67.5^\circ$ & $135^\circ$ & $22.5^\circ$\\
  \hline
\end{tabular}}
\caption{The chosen unitary gates and corresponding setting angles of QWPs and HWPs for preparing these gates in our experiment.}
\label{table:gatesetting}
\end{table*}

In this game, every strategy that uses the two devices in the same input-output direction (either forward direction for both devices or backward direction for both devices) will fail at least 11\% of the times \cite{chiribella2022quantum}.   This lower bound applies even if the relative order between the two gates is indefinite:   for example, combining the forward processes $U$ and $V$ into the coherently controlled gate $UV  \otimes |0\>\<0|  +  VU \otimes |1\>\<1|$, an operation known as the quantum switch \cite{chiribella2009beyond,chiribella2013quantum}, does not offer any reduction of the error probability. More generally, no coherently controlled choice of quantum circuits containing a single use of the forward processes  $(U, V)$, or a single use of the backward processes $(U^T, V^T$), can reduce the error probability below $11\%$.  The same bound applies also to all operations that combine the forward processes  (backward processes)  in a general form of indefinite order \cite{oreshkov2012quantum,chiribella2013quantum}.   

In contrast, a player that uses the two gates in a superposition of the opposite input-output directions $(U,  V^T)$ and $(U^T,  V)$, can bring the error probability arbitrarily close to zero. This result is achieved by combining the two devices into a controlled unitary gate    $W  =  UV^T \otimes |0\>\<0| + U^TV \otimes |1\>\<1|$. Then, the relations  $U^TV = UV^T$ and  $U^TV = -UV^T$ guarantee that one has either $W =  UV^T\otimes I$ or $W   =  UV^T\otimes Z$, respectively, with $Z  =  |0\>\<0|  -|1\>\<1|$. If the control qubit is initialized  in the   state $|+\> := (|0\> + |1\>)/\sqrt 2$, the gates  $UV^T\otimes I$ or $W   =  UV^T\otimes Z$ turn it to the orthogonal states $|+\>$ and $|-\>  :  =  (|0\>  -  |1\>)/\sqrt 2$, respectively. Hence, a projective measurement of the control system can determine without error which of the two alternative relations holds.

\begin{figure*}[htbp]
    \centering
    \includegraphics[width=0.75\linewidth]{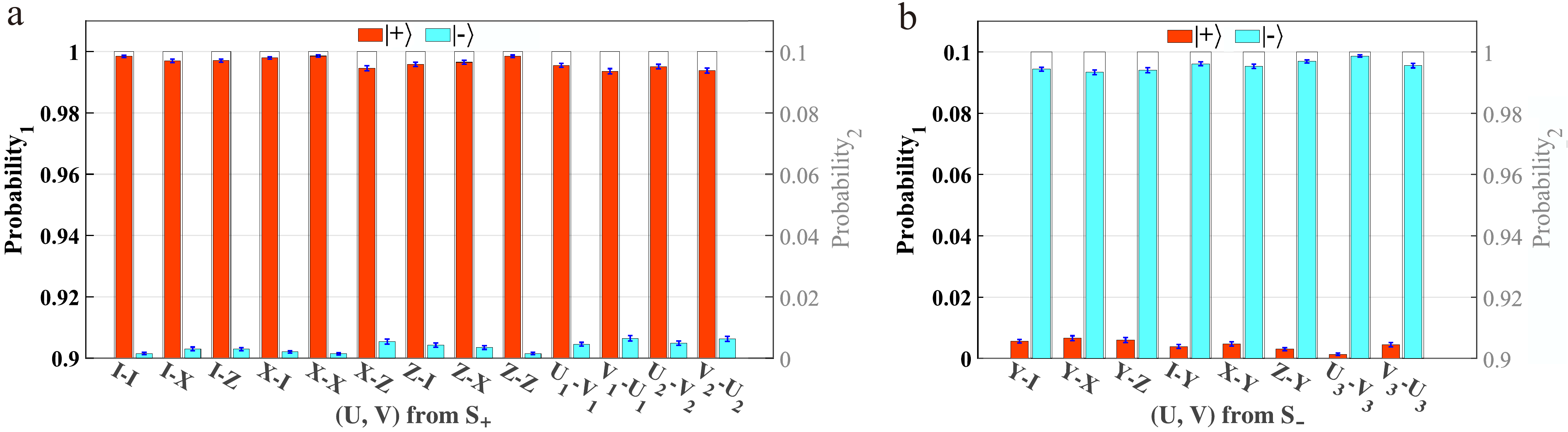}
    \caption{\emph{Experimental data for the quantum game.}  The figure shows the experimentally estimated probabilities of the outcomes $+$ (red bars) and $-$ (cyan bars) of the Pauli-$X$ measurement used to distinguish two alternative properties of an unknown gate pair $(U,V)$. The measurement data refers to 21  different gate pairs, corresponding to different combinations of single-qubit gates in the set $\{  I,  X,  Y,  Z,  U_1, U_2,  U_3, V_1, V_2,  V_3\}$. 
    Gate pairs in the set $\set S_+$ ($\set S_-$) satisfy the condition  $UV^T  =  U^T V$  ($UV^T  = - U^T V$). The experimental data show that the outcome +  ($-$) has high probability to occur for gate pairs in the $\set S_+$  ($\set S_-$), offering a near unit winning probability in the game. }
    \label{fig:result1}
\end{figure*}

The experimental data from our implementation are provided in Fig.  \ref{fig:result1}. The average success probability over all pairs of gates is  $99.6 \pm 0.18\%$, while  the worst-case error probability over all trials is approximately $0.68\pm 0.19\%$. This value is approximately 16 times smaller than 11\%, the lower bound on the error probability for all possible strategies with definite-input-output direction.

Before concluding this section, we provide a brief discussion on the roles of the Mach-Zehnder visibility and single-photon purity. Imperfect  Mach-Zehnder visibility is the main source of errors in our seup. Ideally, if the visibility were perfect,  the photons' output port would be fully  determined by the commutation properties of the unitary gates tested in the game.  With a realistic interferometer, however, some of the photons will reach the wrong port,  leading the player to give a wrong answer.  The probability that the player answers correctly, per detector click,  is then given by $P_{\rm right}  =  n_{\rm right}/(n_{\rm right}+  n_{\rm wrong})$, where $n_{\rm right}$  ($n_{\rm wrong}$) is the number of clicks at  the right (wrong) detector.  The probability of success in the game is then related to the Mach-Zehnder visibility $V= ( n_{\rm right}-  n_{\rm wrong})/(n_{\rm right}+  n_{\rm wrong})$ via the relation $P_{\rm succ}  =  (1+V)/2$. In our experiment,  the visibility is $V=0.9921\pm0.0035$, which yields the value $P_{\rm succ}  = (1+V)/2  =  0.9960\pm0.0018$.

In the above analysis, we considered the success probability per detector click. Another relevant quantity is the success probability per photon entering our setup.    
Ignoring  photon loss, detector efficiency, and detector noise,  the total number of detector clicks $n_{\rm click}  =  n_{\rm right} +  n_{\rm wrong}$ in our experiment is close to the number  of photons $n_{\rm photon}$ entering into the setup, up to a small  correction due to the presence of multiphotons generated by SPDC.  SPDC  generates  pairs of $k$-photons with probability $p_k$, approximately proportional to $\left(P_{\rm pump}\right)^{k}$ where  $P_{\rm pump}$ is the pump power.  A $k$-photon, however, is counted  as a single event by the PerkinElmer  detectors. 
 This implies that, when $n_{\rm photon}$ is large, there will be approximately  $n_{\rm photon} \, p_k/k$ clicks originating from the $k$-photon contribution of SPDC.  Hence, the total number of clicks at the output of the Mach-Zehnder interferometer will be $n_{\rm click}    =  n_{\rm photon} (p_1  +  p_2/2  + p_3/3  + \dots)$. In practice,  in the low power regime of our experiment ($P_{\rm pump}  \approx  2.5~mW$),   $p_k$ is negligible for $k>2$, and therefore one can assume $p_1+  p_2  \approx 1$.  
 
 In fact, the correction due to the $p_2$ term is itself almost negligible.  The value of $p_2$ is closely related to the single-photon purity, given by $\sqrt{ 1  - g^{(2)} (0)}$ where $g^{(2)} (0)$ is the heralded idler-idler self-correlation function.  The latter can be estimated from experimental data according to a heralded Hanbury-Brown-Twiss configuration \cite{guo2017parametric}.  In this configuration,  a SPDC process generates a pair of photons, with the signal photon acting as a herald (detected by a detector H) and the idler photon  split into two path modes by a 50/50 beamsplitter, with detectors A and B at its output ports. The self-correlation function is then given by 
$g^{(2) } (0)=(n_{\rm H} n_{\rm HAB})/(n_{\rm HA} n_{\rm HB})$,
where $n_{\rm H}$ is the number of counts at  detector H, while  $n_{\rm HAB}$, $n_{\rm HA}$, and $n_{\rm HB}$ are the coincidence rates at zero delay time at detectors H/A/B, H/A, and H/B, respectively.  In our experiment,  $n_{\rm HAB}$  can be estimated as $n_{\rm HAB}=2\tau \eta n_{\rm A} n_{\rm B}$, where $\tau$ is the coincidence window,  $\eta$ is the coincidence efficiency of the SPDC source, and $n_{\rm A}$ and  $n_{\rm B}$ are the counts at detectors A and B, respectively.  The coincidence rate  was 20000, $n_H$ and $n_A+n_B$ were 61000 and 58000, respectively, and $\tau$ was set to be 1 ns.  Using these data, we can estimate $\eta$, $n_{\rm A}$, $n_{\rm B}$, $n_{\rm HA}$, $n_{\rm HB}$, and $n_{\rm HAB}$ to be 0.33, 29000, 29000, 10000, 10000, and 0.56, respectively. From these value, we then obtain $g^{(2) } (0)\approx 0.0003$, corresponding to a single-photon purity $\sqrt{ 1  - g^{(2) } (0) } \approx 0.9998$. 

From the coincidence rates, one can also directly estimate the two-photon pair event probability, which is given by
$p_2  =  (2n_{\rm HAB})/(n_{\rm HA}+n_{\rm HB} )\approx 6  \times  10^{-5}$.
 This value implies that in our game  the success probability per photon is    $P_{\rm right}^{\rm photon} = P_{\rm succ} \,  n_{\rm click}/n_{\rm photon}  \approx  P_{\rm right} \,  [p_1  +  p_2/2]  \approx  P_{\rm right} \,  [1-  p_2/2] \approx 0.9960$. Note that the success probability per photon is equal, within the experimental error bars, to the success probability per click $P_{\rm succ} =  0.9960\pm0.0018$, as a result of the high single-photon purity in our experiment.


\section{A strengthened advantage  in the quantum game}\label{app:strengthenedgame}

Here we show that the advantage demonstrated in our experiment holds even if our setup is compared with arbitrary setups that use the control-unitary gates ${\tt ctrl}  - U  =  I \otimes |0\>\<0|  +  U \otimes |1\>\<1|$ and ${\tt ctrl}  - V  =  I \otimes |0\>\<0|  +  V \otimes |1\>\<1|$ instead of the original gates $U$ and $V$.   As long as these controlled gates are used in a fixed input-output direction (either the forward direction for both gates or the backward direction for both gates), the error probability in the game cannot go below 5.6\%.   Notably, this result holds even if the relative order between the gates  ${\tt ctrl}  - U$  and ${\tt ctrl}  - V$ is indefinite.  
  
 To derive this result,  we adapt the method developed in Ref. \cite{chiribella2022quantum} to find the minimal worst-case error over all strategies using the gates $U$  and $V$ in a fixed input-output direction, equal for both gates. Here we replace the 2-by-2 matrices $U$ and $V$ with their controlled versions ${\tt ctrl}  - U$  and ${\tt ctrl}  - V$,   mathematically equivalent to  the block diagonal matrices  
 $\widetilde U=  \left(
 \begin{array}{c|c}   
 I & {\bf 0}\\
 \hline
 {\bf 0}   &   U  
 \end{array}
 \right)$
and 
$\widetilde V  = \left(
 \begin{array}{c|c}   
 I & {\bf 0}\\
 \hline 
 {\bf 0}   &   V  
 \end{array}
 \right)$, respectively. 
With this substitution, the minimum probability of error over all possible strategies with definite input-output direction can be cast in the SDP form 
\begin{alignat}{2}
    \label{eq:gamesdp}
    & \text{minimize} \quad p_{\rm err} &&\\
    & \text{subject to}   &&\\
     &\bb \widetilde U|_{A_{\rm O} A_{\rm I}}  \bb \widetilde V|_{B_{\rm O} B_{\rm I}} ~ P~~   |\widetilde U\kk_{A_{\rm O} A_{\rm I}} | \widetilde V\kk_{B_{\rm O} B_{\rm I}}  \geq 1 - p_{\rm err}  && \nonumber \\ 
    & \forall (U, V) \in \set S_+ \, ,  && \nonumber \\
    &  \bb \widetilde U|_{A_{\rm O} A_{\rm I}}  \bb \widetilde V|_{B_{\rm O} B_{\rm I}} ~ P~~   |\widetilde U\kk_{A_{\rm O} A_{\rm I}} | \widetilde V\kk_{B_{\rm O} B_{\rm I}}  \leq p_{\rm err} &&\nonumber \\ 
    & \forall (U, V) \in \set S_- \, ,  && \nonumber \\
    & 0 \leq P \leq S \nonumber \, ,  && \\
    &  S \in  \set{Det}   (A_{\rm I} ,A_{\rm O};  B_{\rm I}, B_{\rm O})\, .  &&
\end{alignat}
The optimization runs over the variables $p_{\rm err}$,   $P$, and $S$, where  $p_{\rm err}$ is the probability of error, $P  \in  L(  \spc H_{A_{\rm O}} \otimes \spc H_{A_{\rm I}}  \otimes   \spc H_{B_{\rm O}} \otimes \spc H_{B_{\rm I}})$ is a positive operator, and $S$  is an operator in the set $\set{Det}   (A_{\rm I} ,A_{\rm O};  B_{\rm I} ,B_{\rm O})$, consisting of all operators $S$ satisfying the following constraints: 
\begin{align}
    \nonumber S \geq 0 \, , \\
    \nonumber {}_{[A_IA_O]}S = {}_{[A_IA_OB_O]}S \, ,\\
    \nonumber {}_{[B_IB_O]}S = {}_{[A_OB_IB_O]}S \, ,\\
    S = {}_{[A_O]}S + {}_{[B_O]}S - {}_{[A_OB_O]}S \, .
\end{align}  

Operationally, the pair  $\{ P, S-P \}$ represents a binary-outcome test that probes the gates $\widetilde U$ and $\widetilde V$ in a definite input-output direction and possibly in an indefinite order relative to one another \cite{chiribella2019quantum,bavaresco2021strict}. The above SDP yields the minimum error probability achievable by this type of setup.     Numerical computation with the Python-embedded modeling language CVXPY \cite{diamond2016cvxpy,agrawal2018rewriting} yields the results that the minimal $p_{\rm err}$ is 5.6\%. The same result holds if we replace the forward gates  ${\tt ctrl}  - U$  and ${\tt ctrl}  - V$ with the backward gates  ${\tt ctrl}  - U^T$  and ${\tt ctrl}  - V^T$.    Hence, a player that uses both gates in the same input-output direction will always have a finite error probability,  even if the player has access to controlled unitary gates.


\end{document}